\documentclass[11pt, a4paper]{article}
\pdfoutput=1 

\usepackage[height=22.5cm,width=17.5cm]{geometry}

\usepackage{graphicx,rotating}     
\usepackage[bookmarksopen,colorlinks=true,linkcolor=teal,
citecolor=darkgoldenrod,urlcolor=darkgoldenrod,linktoc=all]{hyperref}

\usepackage{amsmath}
\usepackage{mathtools}
\usepackage{amsfonts}
\usepackage{amssymb}
\usepackage{braket}
\usepackage{bm}
\usepackage{cite}
\usepackage{comment}
\usepackage{xcolor}
\usepackage[utf8]{inputenc}
\usepackage[footnotesize]{caption}
\usepackage{bbold}

\newcommand{\abs}[1]{\left\vert {#1} \right\vert}

\makeatletter
\@addtoreset{equation}{section}

\makeatother

\usepackage{multicol}
\definecolor{indianred}{HTML}{CD5C5C}
\definecolor{teal}{HTML}{008080}
\definecolor{forestgreen}{HTML}{228B22}
\definecolor{darkgoldenrod}{HTML}{B8860B}

\begin{document}

\hypersetup{pageanchor=false}
\begin{titlepage}

\begin{center}

\hfill DESY 21-027\\

\vskip 1.in

{\bfseries
{ \huge 
Preheating from target space curvature\vspace{5mm} \\ 
and unitarity violation:\vspace{6mm}} \\
{\LARGE Analysis in field space}} \\
\vskip .8in

{\Large Yohei Ema$^{a}$, Ryusuke Jinno$^{a}$, Kazunori Nakayama$^{b,c}$, Jorinde van de Vis$^{a}$}

\vskip .3in
\begin{tabular}{ll}
$^a$& \!\!\!\!\!\emph{DESY, Notkestra{\ss}e 85, D-22607 Hamburg, Germany}\\
$^b$& \!\!\!\!\!\emph{Department of Physics, Faculty of Science, The University of Tokyo,}\\[-.15em]
& \!\!\!\!\!\emph{Bunkyo-ku, Tokyo 113-0033, Japan}\\
$^c$& \!\!\!\!\!\emph{Kavli IPMU (WPI), The University of Tokyo, Kashiwa, Chiba 277-8583, Japan}
\end{tabular}

\end{center}
\vskip .6in

\begin{abstract}
\noindent
We study particle production and unitarity violation caused by a curved target space right after inflation.
We use the inflaton field value instead of cosmic time as the time variable,
and derive a semiclassical formula for the spectrum of produced particles.
We then derive a simple condition for unitarity violation during preheating,
which we confirm by our semiclassical method and numerical solution.
This condition depends not only on the target space curvature 
but also on the height of the inflaton potential at the end of inflation.
This condition tells us, for instance, 
that unitarity is violated in running kinetic inflation and Higgs inflation,
while unitarity is conserved in $\alpha$-attractor inflation and Higgs-Palatini inflation.
\end{abstract}

\end{titlepage}

\tableofcontents
\renewcommand{\thepage}{\arabic{page}}
\renewcommand{\thefootnote}{$\natural$\arabic{footnote}}
\setcounter{footnote}{0}
\hypersetup{pageanchor=true}

\section{Introduction}


Inflation is a successful paradigm that describes the beginning of the universe
and provides seeds for large scale structure
and the cosmic microwave background (CMB) anisotropies.
Inflation is typically driven by the potential energy of a scalar field, 
\emph{i.e.}, an inflaton, that slowly rolls down its potential.
After inflation, the inflaton typically oscillates around the minimum of its potential
and its energy is released by particle production,
a process referred to as (p)reheating.
The (p)reheating dynamics is of great importance since the inflationary observables such as 
the spectral index and tensor-to-scalar ratio depend on the reheating temperature
through the duration of the reheating phase.
In addition, interesting physical processes in the early universe such as
baryogenesis and dark matter production often depend on the reheating temperature.

Recently, it was pointed out that the preheating dynamics of Higgs inflation
can be more violent than previously thought~\cite{Ema:2016dny}. 
In Higgs inflation the standard model Higgs is identified as the inflaton and,
with the help of a nonminimal coupling to gravity $\xi$, the model
is consistent with the CMB observation~\cite{Futamase:1987ua,CervantesCota:1995tz,Bezrukov:2007ep,Akrami:2018odb}, 
making it one of the most popular inflation models on the market.
Ref.~\cite{Ema:2016dny} found that an effective mass term of the Goldstone (or equivalently longitudinal gauge) bosons 
shows a ``spiky" feature as the Higgs passes the origin.
The typical energy scale of particles produced by this spike exceeds the cut-off scale of the theory for 
$\xi \gtrsim \mathcal{O}(100)$,\footnote{
	The CMB requires $\xi^2 \simeq 2\times 10^9 \lambda$,
	where $\lambda$ is the Higgs quartic coupling,
	and hence this is the case unless $\lambda$ is 
	(tuned to be) tiny at the inflationary energy scale.
} resulting in unitarity violation. 
In other words, this phenomena invalidates the low energy description of Higgs inflation 
and requires a UV completion.\footnote{
	We discuss different viewpoints on the unitarity issue of Higgs inflation in Sec.~\ref{sec:specmodels}.
}
The unitarity violation of Higgs inflation was originally found 
by studying the dynamics of the inflaton in the Jordan frame, 
but this is also nicely understood as a result of the target space curvature in the Einstein frame~\cite{DeCross:2016cbs,Sfakianakis:2018lzf}.
Since there are many other inflation models that have nontrivial target space curvatures in the Einstein frame,
it is then natural to ask if unitarity violation happens in those other models.

In this paper, we study particle production and unitarity violation 
from the target space curvature during preheating in detail. Throughout this work, we study preheating by a set of linearized mode equations, and we neglect the backreaction to the inflaton field. This treatment allows us to obtain analytical expressions for the spectrum after the first inflaton zero crossing.\footnote{In the case of efficient particle production, the linearized treatment quickly breaks down and one should resort to lattice simulations for an accurate description of the particle production process. Generally speaking, the linearized equations give a good description of the spectrum after the first inflaton zero crossing. Even when preheating is very strong, it allows us to estimate the typical momentum of the produced particles.}
In Sec.~\ref{sec:pp_field_space}, we study particle production of a scalar field~$\chi$ 
coupled to the inflaton~$\phi$.\footnote{
	Since the target space is trivial if there is only one scalar field,
	we always assume that there are (at least) two scalar fields in this paper.
	Remember that, for instance, Higgs inflation has four real scalar degrees of freedom.
}
This section is intended to be generic,
and hence we take the effective mass term of~$\chi$ to be a general
function of~$\phi$ and its velocity $\dot{\phi}$ without specifying its origin.
In general, it is challenging to study particle production analytically when the kinetic term of the inflaton is nontrivial,
since the inflaton is not a simple function of the cosmic time $t$ in such a case.
Instead, we use the inflaton field value $\phi$ itself as the time variable and work solely in the scalar field space,
not referring to $t$. 
In this method, the inflaton velocity is expressed as a function of $\phi$ by using
an (approximately) conserved quantity such as the inflaton energy density.
This enables us to apply a phase integral approximation 
to obtain a semiclassical formula of the spectrum that reproduces the numerical results quite well.

We then address the question of unitarity violation in Sec.~\ref{sec:preheating_unitarity}.
In particular, we find that the following simple condition works as a criterion for unitarity violation:
\begin{align}
	V(\Phi) \gtrsim \Lambda^4,
	\label{eq:unitarity_intro}
\end{align}
where $V(\Phi)$ is the height of the inflaton potential at the end of inflation with $\Phi$ the inflaton amplitude,
and $\Lambda$ is the typical mass scale of the target space curvature.
Note that $\Lambda$ is the cut-off scale of the theory at the same time since it enters
the scattering amplitude of, \emph{e.g.}, $\phi \phi \rightarrow \chi \chi$.
For instance, $V(\Phi) \sim \lambda M_P^4/\xi^2$ and $\Lambda \sim M_P/\xi$
for Higgs inflation with $\lambda$ the Higgs quartic coupling and $M_P$ the reduced Planck scale~\cite{Burgess:2009ea,Barbon:2009ya,Burgess:2010zq,Hertzberg:2010dc,Bezrukov:2010jz},
and hence the condition~\eqref{eq:unitarity_intro} reads
\begin{align}
	\lambda \xi^2 \gtrsim 1,
\end{align}
which agrees with the condition known in the literature~\cite{Ema:2016dny}.
The condition~\eqref{eq:unitarity_intro} tells us that not only the target space curvature but also the height
of the inflaton potential is important for unitarity violation.
We derive this condition based on an intuitive argument,
and confirm it by the method developed in Sec.~\ref{sec:pp_field_space} 
as well as numerical computations.
We find in particular that this condition explains the different character of the preheating dynamics
of Higgs and Higgs-Palatini inflation~\cite{Bauer:2008zj,Bauer:2010jg,Rubio:2019ypq}.
We also point out that unitarity violation can occur in running kinetic inflation~\cite{Takahashi:2010ky,Nakayama:2010kt}, 
which, to our knowledge, has not yet been noted in the literature (except for a brief comment in Ref.~\cite{Ema:2016dny}).
In addition, we see that the method in Sec.~\ref{sec:pp_field_space} provides a good approximation of the spectrum
even when unitarity is preserved.

Although our main focus is on the preheating dynamics after inflation,
particle production and unitarity violation from the target space curvature are possible even beyond this context.
As an example, we comment on a supersymmetric axion model in the end of Sec.~\ref{sec:preheating_unitarity}.
This model acquires a nontrivial target space as an induced metric after integrating our heavy degrees of freedom.
The radial component of the Peccei-Quinn field, or the saxion field, plays the role of $\phi$,
and its motion induces a spiky feature in the effective mass term of the axion $\chi$ that may cause unitarity violation
within the low energy description.

We finally summarize our results in Sec.~\ref{sec:summary},
with comments on possible UV completions of running kinetic inflation and Higgs inflation.
 
\section{Semiclassical analysis of particle production in field space}
\label{sec:pp_field_space}

In this section, we study the particle production of a scalar particle $\chi$
during an inflaton oscillation epoch after inflation.
In particular, we use the inflaton field value instead of the cosmic time as our time variable,
which we may refer to as an analysis in the (scalar) field space.
This enables us to derive an semiclassical expression of the occupation number
after one inflaton oscillation.
We will show that the semiclassical analysis reproduces numerical results well
in two specific cases.
These results turn out to be useful to understand particle production from a target space curvature,
as we will see in Sec.~\ref{sec:preheating_unitarity}. 
We ignore the Hubble expansion in this section 
since it is irrelevant for our discussion.

\subsection{Preliminary}
In this subsection, we summarize the basic equations for particle production.
We consider the following Lagrangian for the inflaton field $\phi$:
\begin{align}
	\mathcal{L} &= \frac{1}{2}h_{\phi\phi}\left(\partial \phi\right)^2 - V,
	\label{eq:lagrangian_inflaton}
\end{align}
where $h_{\phi\phi} = h_{\phi\phi}(\phi)$ and $V = V(\phi)$ are general functions of $\phi$.
We will see in Sec.~\ref{sec:preheating_unitarity} that this form of Lagrangian describes, \emph{e.g.},
running kinetic inflation~\cite{Takahashi:2010ky,Nakayama:2010kt}, 
Higgs inflation~\cite{Futamase:1987ua,CervantesCota:1995tz,Bezrukov:2007ep}, 
$\alpha$-attractor inflation~\cite{Kallosh:2013hoa,Kallosh:2013yoa} 
and Higgs-Palatini inflation~\cite{Bauer:2008zj,Bauer:2010jg}.\footnote{
	This corresponds to the action in the Einstein frame for (Palatini) Higgs inflation.
} The inflaton equation of motion is given by
\begin{align}
	&h_{\phi\phi} \ddot{\phi} + \frac{1}{2}h_{\phi\phi}' \dot{\phi}^2 + V' = 0,
	\label{eq:eom_inflaton_t}
\end{align}
where the dot and prime denote the derivatives with respect to $t$ and $\phi$, respectively.
This system has a conserved quantity,
\begin{align}
	\frac{1}{2}h_{\phi\phi}\left(\phi\right)\dot{\phi}^2 + V\left(\phi\right) = V\left(\Phi\right),
	\label{eq:conserved_quantity_general}
\end{align}
where $\Phi$ is the initial inflaton amplitude.

We consider the production of a scalar particle $\chi$ whose action is given by
\begin{align}
	\mathcal{L}_\chi = \frac{1}{2}\left(\partial\chi\right)^2
	- \frac{1}{2}m_\chi^2(\phi, \dot{\phi})\chi^2.
\end{align}
Here we do not specify the origin of the mass $m_\chi^2$ and 
take it as a general function of $\phi$ and $\dot{\phi}$.
We will study the mass originating from the curvature of the scalar field target space
in Sec.~\ref{sec:preheating_unitarity}.
Moving to Fourier space, we obtain the mode equation as
\begin{align}
	\ddot{\chi}_k + \left(k^2 +m_\chi^2\right)\chi_k = 0.
\end{align}
We impose plane wave initial conditions
\begin{align}
	\chi_k = \frac{1}{\sqrt{2\omega_k}},
	\quad
	{\dot{\chi}}_k = -i\sqrt{\frac{\omega_k}{2}},
	\label{eq:init_cond_chi}
\end{align}
where the frequency is given by
\begin{align}
	\omega_k^2 = k^2 + m_\chi^2.\label{eq:mode}
\end{align}

Since the mass term and hence the frequency depend on time through $\phi$ and $\dot{\phi}$,
positive and negative frequency modes get mixed with each other as time evolves,
which is interpreted as particle production (see, \emph{e.g.}, Ref.~\cite{Birrell:1982ix}).
This mixing is described by the Bogoliubov coefficients $\alpha_k$ and $\beta_k$
whose equations of motion are given by
\begin{align}
	\dot{\alpha}_k &= \frac{1}{4\omega_k^2}\frac{d \omega_k^2}{dt} \beta_k e^{2i\int^t d t\, \omega_k},
	\quad
	\dot{\beta}_k = \frac{1}{4\omega_k^2}\frac{d \omega_k^2}{dt} \alpha_k e^{-2i\int^t dt\, \omega_k}.
	\label{eq:alpha_beta_t}
\end{align}
The initial condition~\eqref{eq:init_cond_chi} translates to
\begin{align}
	\alpha_k = 1,
	\quad
	\beta_k = 0,
\end{align}
at the initial time.
Finally, the occupation number of a given mode is computed as
\begin{align}
	f_k = \abs{\beta_k}^2.
\end{align}

\subsection{Inflaton field value as a time variable}

If $h_{\phi\phi}$ is nontrivial,
the dynamics of the inflaton $\phi$ generally depend on $t$ in a complicated manner,
and an analytic expression can be obtained only in a some specific cases and/or for a limited field range.
This makes an analytical estimation of the particle production quite difficult
(see, \emph{e.g.}, Ref.~\cite{Ema:2016dny} in the context of Higgs inflation).

In this paper, we propose using the inflaton field value $\phi$
instead of $t$ as the time variable
to study the particle production within one oscillation.
The inflaton field value $\phi$ is a monotonic function of $t$ within half of an oscillation, 
and hence it is equally a good time variable.
The inflaton velocity $\dot{\phi}$ can be expressed as a function of $\phi$
by exploiting the conserved quantity~\eqref{eq:conserved_quantity_general}.
A virtue of this method is that, once $h_{\phi\phi}$ and $V$ are given, it is straightforward
to derive an analytic formula for $m_\chi^2$ in terms of $\phi$ without any approximation, 
as we will see explicitly below.
This enables us to perform a semiclassical analysis of the particle production 
and estimate the spectrum analytically.

In terms of $\phi$ instead of $t$, the Bogoliubov coefficients satisfy
\begin{align}
	\frac{d{\alpha}_k}{d\phi} &= \frac{1}{4\omega_k^2}\frac{d \omega_k^2}{d \phi} \beta_k 
	\exp\left(2i\int^\phi d\phi\, \frac{\omega_k}{\dot{\phi}}\right),
	\quad
	\frac{d{\beta}_k}{d\phi} = \frac{1}{4\omega_k^2}\frac{d \omega_k^2}{d\phi} \alpha_k 
	\exp\left(-2i\int^\phi d\phi\, \frac{\omega_k}{\dot{\phi}}\right),
\end{align}
where $\omega_k$ is now understood as a function of $\phi$.
The exponent contains the velocity $\dot{\phi}$ due to a change of integration variable.
It follows from Eq.~\eqref{eq:conserved_quantity_general} that 
$\dot{\phi}$ is expressed as a function of $\phi$ as
\begin{align}
	\dot{\phi}^2 = 2\frac{V\left(\Phi\right) - V\left(\phi\right)}{h_{\phi\phi}\left(\phi\right)}.
\end{align}
Since $\dot{\phi}$ has a definite sign within half of an oscillation,
there is no issue in solving the square in this equation.
We define $\Phi$ as the inflaton amplitude at the end of inflation. Focusing on particle production \emph{after} inflation and ignoring particle production \emph{during} inflation,
we take the initial condition as
\begin{align}
	\alpha_k\left(\Phi\right) = 1,
	\quad
	\beta_k\left(\Phi\right) = 0,
\end{align}
and solve the equations until $\phi = -\Phi$. 
The occupation number after the first zero crossing is then given by
\begin{align}
	f_k = \abs{\beta_k\left(-\Phi\right)}^2.
\end{align}

Here is a comment. Although we restrict ourselves to Eq.~\eqref{eq:lagrangian_inflaton} in this paper,
it is obvious that the idea of using $\phi$ instead of $t$ as a time variable
can be applied for a broader class of models,
such as the generalized Galileon theory~\cite{Deffayet:2011gz,Kobayashi:2011nu}.
One caution here is that it is sometimes nontrivial to find an approximately conserved quantity.
The energy density is approximately conserved even if one turns on the cosmic expansion 
in the model~\eqref{eq:lagrangian_inflaton},
and hence one can use it to express $\dot{\phi}$ by $\phi$ as long as the process of one's interest
occurs faster than the expansion as we will see in Sec.~\ref{sec:preheating_unitarity}.
For a more general model, however, the energy density can highly oscillate, and cannot be used 
to express $\dot{\phi}$ in terms of $\phi$ (see, \emph{e.g.}, Ref.~\cite{Jinno:2013fka,Ema:2015oaa}).
In such a case, an adiabatic invariant discussed in Ref.~\cite{Ema:2015eqa} will be a useful alternative to the energy density.

\subsection{Semiclassical analysis}

We now explain our semiclassical analysis to estimate the occupation number.
We exploit the Born approximation, $\alpha_k \simeq 1$, 
that is valid when the occupation number is small, which results in
\begin{align}
	\frac{d{\beta}_k}{d\phi} \simeq \frac{1}{4\omega_k^2}\frac{d \omega_k^2}{d\phi} 
	\exp\left(-2i\int^\phi d\phi\, \frac{\omega_k}{\dot{\phi}}\right).
\end{align}
This equation is trivially integrated as
\begin{align}
	f_k \simeq \abs{\int_{-\Phi}^{\Phi} \frac{d\phi}{4}\frac{1}{\omega_k^2}\frac{d \omega_k^2}{d\phi}
	\exp\left(-2i\int^\phi d\phi \frac{\omega_k}{\dot{\phi}}\right)}^2.
	\label{eq:born_general_finitePhi}
\end{align}
We further assume that the inflaton amplitude $\Phi$ is much larger than the field value region
in which the particle production dominantly happens.
Then we can take the limit $\Phi \rightarrow \infty$ and we obtain
\begin{align}
	f_k \simeq \abs{\int_{-\infty}^{\infty} \frac{d\phi}{4}\frac{1}{\omega_k^2}\frac{d \omega_k^2}{d\phi}
	\exp\left(-2i\int^\phi d\phi \frac{\omega_k}{\dot{\phi}}\right)}^2.
	\label{eq:born_general}
\end{align}
\begin{figure}[t]
	\centering
 	\includegraphics[width=0.45\linewidth]{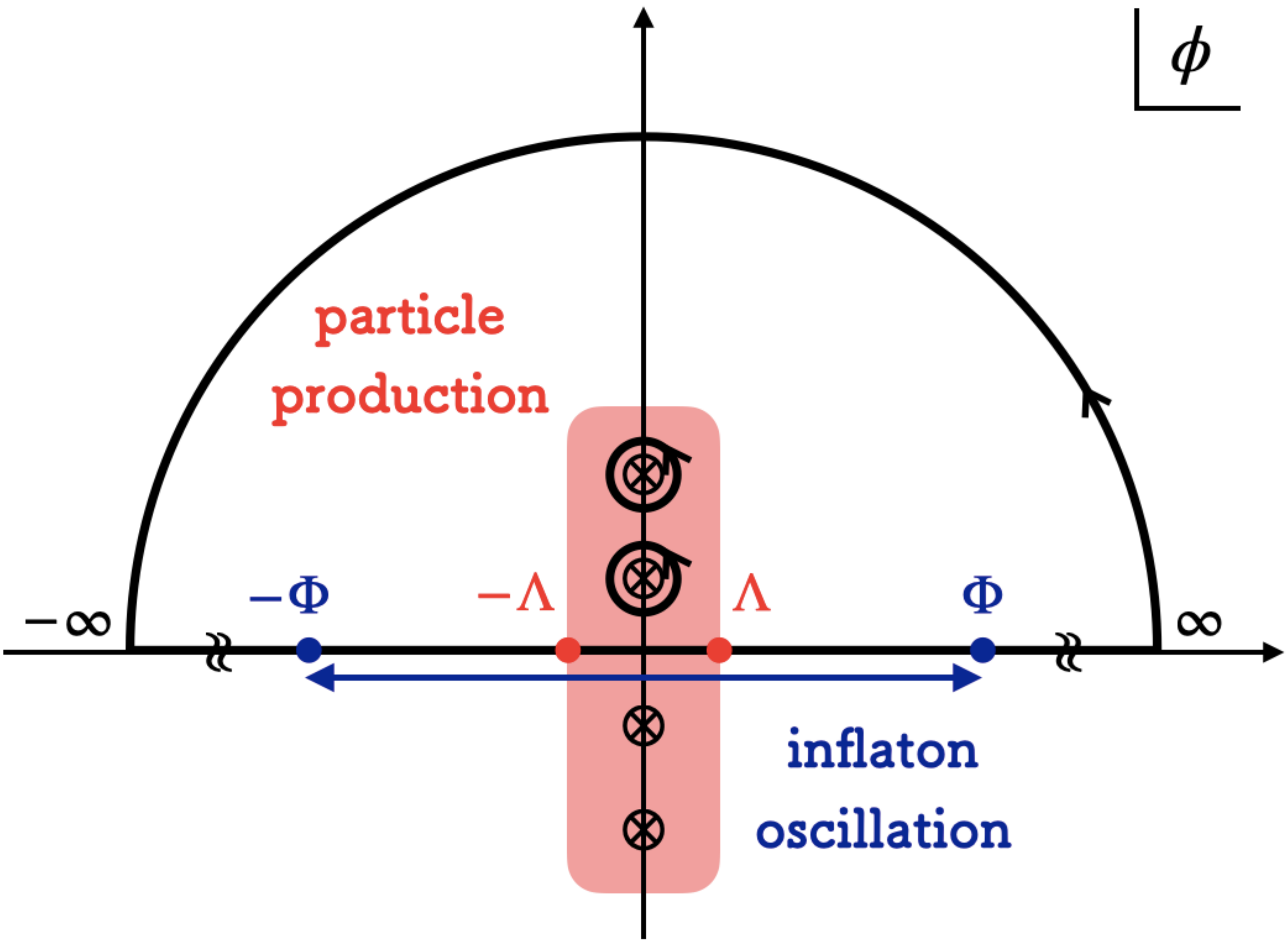}
 	\includegraphics[width=0.45\linewidth]{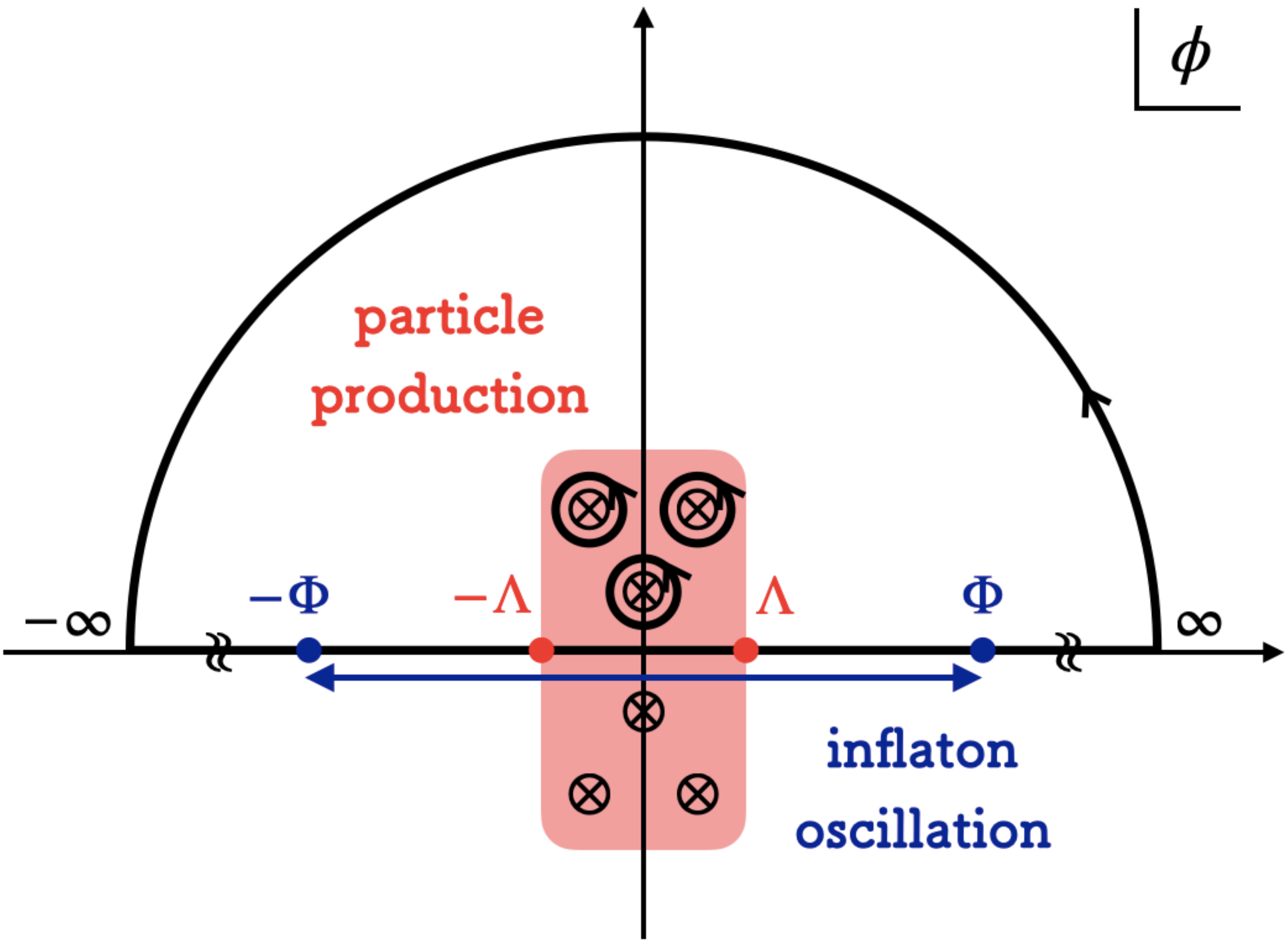}
	\caption{\small Integration contours in the complex $\phi$-plane,
	for two different types of pole structure,
	corresponding to the two examples in Sec.~\ref{subsec:pia_examples}.
	The extension of the integration range from $[-\Phi, \Phi]$ to $[-\infty, \infty]$ 
	is justified when the particle production dominantly occurs only around the origin
	$\abs{\phi} \lesssim \Lambda~(\ll \Phi)$.
	The extended integration contour can then be closed at complex infinity,
	reducing the integral to the sum of the residues at the poles.}
	\label{fig:schematic_semiclassical}
\end{figure}

We now rely on the phase integral method (see, \emph{e.g.}, Refs.~\cite{Dumlu:2010ua, Dumlu:2011rr} and references therein).
We assume that the integral contour of Eq.~\eqref{eq:born_general} 
can be closed either in the upper or lower half of the complex $\phi$-plane.
This is typically the case due to the exponential factor in Eq.~\eqref{eq:born_general}.
Then, according to Cauchy's residue theorem, the integral can be replaced by a sum of poles
\begin{align}
	f_k \simeq
	\frac{\pi^2}{4}\abs{\sum_{\phi_\otimes}
	\mathrm{Res}_{\phi = \phi_\otimes}\left[\frac{1}{\omega_k^2}\frac{d\omega_k^2}{d\phi}
	\exp\left(-2i\int^{\phi}_0 d\phi\,\frac{\omega_k}{\dot{\phi}}\right)\right]}^2,
	\label{eq:pia_general}
\end{align}
where the sum is taken over all the poles $\phi = \phi_\otimes$ in either the upper or lower half planes. 
If the inflaton has a $\mathbb{Z}_2$ symmetry $\phi \leftrightarrow -\phi$, the poles always appear in pairs,
and hence there is no difference between taking either the upper or lower half plane,
as long as one ensures that the spectrum is exponentially suppressed, not enhanced,
in the final expression.
Since the overall phase does not contribute to the spectrum,
we take the lower end of the integral in the exponent as $\phi = 0$ in this formula.
Fig.~\ref{fig:schematic_semiclassical} describes the schematic picture of our procedure outlined above.

Here we clarify our assumptions and limitations of the above formula.
First, we used the Born approximation, and hence the above formula works only when
the occupation number does not exceed unity.
We will see below that this is indeed a good approximation in the case of our interest.
Second, we took the limit $\Phi \rightarrow \infty$ for the end points of the integral.
This limit corresponds to, \emph{e.g.}, $\xi \gg 1$ in Higgs inflation,
which is indeed the case of our main interest.
We will also see in Sec.~\ref{sec:preheating_unitarity} that this approximation is valid
even when $\xi = 10^2$, for which unitarity is preserved.
Third, we implicitly assumed that the modes of our interest satisfy $\omega_k^2 > 0$ 
so that $\alpha_k$ and $\beta_k$ are well-defined.
If a mode becomes tachyonic during an inflaton oscillation, however, one has to smoothly connect
the regions $\omega_k^2 > 0 $ and $\omega_k^2 < 0$ as in Ref.~\cite{Dufaux:2006ee}.
This requires a separate treatment that is beyond the scope of this paper.
In our context, whether this happens depends on the sign of the target space curvature,
and this prevents us from applying the above semiclassical method to, \emph{e.g.}, $\alpha$-attractor inflation.

In the following, we demonstrate how the general procedure we outlined above works in practice with two examples.
These results will be extensively used in Sec.~\ref{sec:preheating_unitarity}.

\subsection{Examples}
\label{subsec:pia_examples}

In the following examples, we take the following metric
\begin{align}
	h_{\phi\phi} = 1 + \frac{\phi^2}{\Lambda^2},
\end{align}
where $\Lambda$ is a suppression scale.
The conserved quantity then determines
\begin{align}
	\dot{\phi}^2 \simeq \frac{2 V(\Phi)}{1 + \phi^2/\Lambda^2},
	\label{eq:rki_dphi_by_phi}
\end{align}
where we further ignored the $V(\phi)$ term in the numerator which is valid in the limit $\Phi/\Lambda \rightarrow \infty$.
In this example, the approximation that lets Eq.~\eqref{eq:born_general} follow from Eq.~\eqref{eq:born_general_finitePhi}
also corresponds to the limit $\Phi/\Lambda \rightarrow \infty$.

\subsubsection*{Example~1: mass with simple poles}

Here we take the effective mass term
\begin{align}
	m_\chi^2 = c \frac{\dot{\phi}^2}{\Lambda^2},
	\label{eq:single_pole_dphi}
\end{align}
where $c$ is a numerical factor.
By using Eq.~\eqref{eq:rki_dphi_by_phi}, this can be rewritten as
\begin{align}
	m_\chi^2 \simeq 2c\frac{V(\Phi)}{\Lambda^2}\frac{1}{1 + \phi^2/\Lambda^2},
\end{align}
which has a pair of simple poles.
The quantities necessary for the semiclassical analysis are given by
\begin{align}
	\frac{1}{\omega_k^2}\frac{d\omega_k^2}{d\bar{\phi}}
	&= -\frac{2\bar{\phi}}{1+\bar{\phi}^2}\frac{1}{\bar{k}^2\left(1+\bar{\phi}^2\right) + 1}, \\
	\frac{\omega_k^2}{\dot{\bar{\phi}}^2} &= c \left(\bar{k}\left(1+\bar{\phi}^2\right) + 1\right),
\end{align}
where we defined the dimensionless parameters
\begin{align}
	\bar{\phi} \equiv \frac{\phi}{\Lambda},
	\quad
	\bar{k}^2 \equiv \frac{1}{2c} \frac{k^2 \Lambda^2}{V(\Phi)}.
\end{align}
The prefactor of the integral has poles at
\begin{align}
	\bar{\phi} = \pm i, 
	\quad
	\pm i \frac{\sqrt{1+\bar{k}^2}}{\bar{k}},
\end{align}
and hence Eq.~\eqref{eq:pia_general} reads
\begin{align}
	f_k \simeq 
	\frac{\pi^2}{4}\abs{
	\exp\left[-\sqrt{c}\left(1+\left(\bar{k}+\frac{1}{\bar{k}}\right)\arctan\left(\bar{k}\right)\right)\right]
	- \exp\left[-\frac{\pi \sqrt{c}}{2}\frac{\bar{k}^2+1}{\bar{k}}\right]
	}^2.
	\label{eq:pia_simple}
\end{align}
\begin{figure}[t]
	\centering
 	\includegraphics[width=0.45\linewidth]{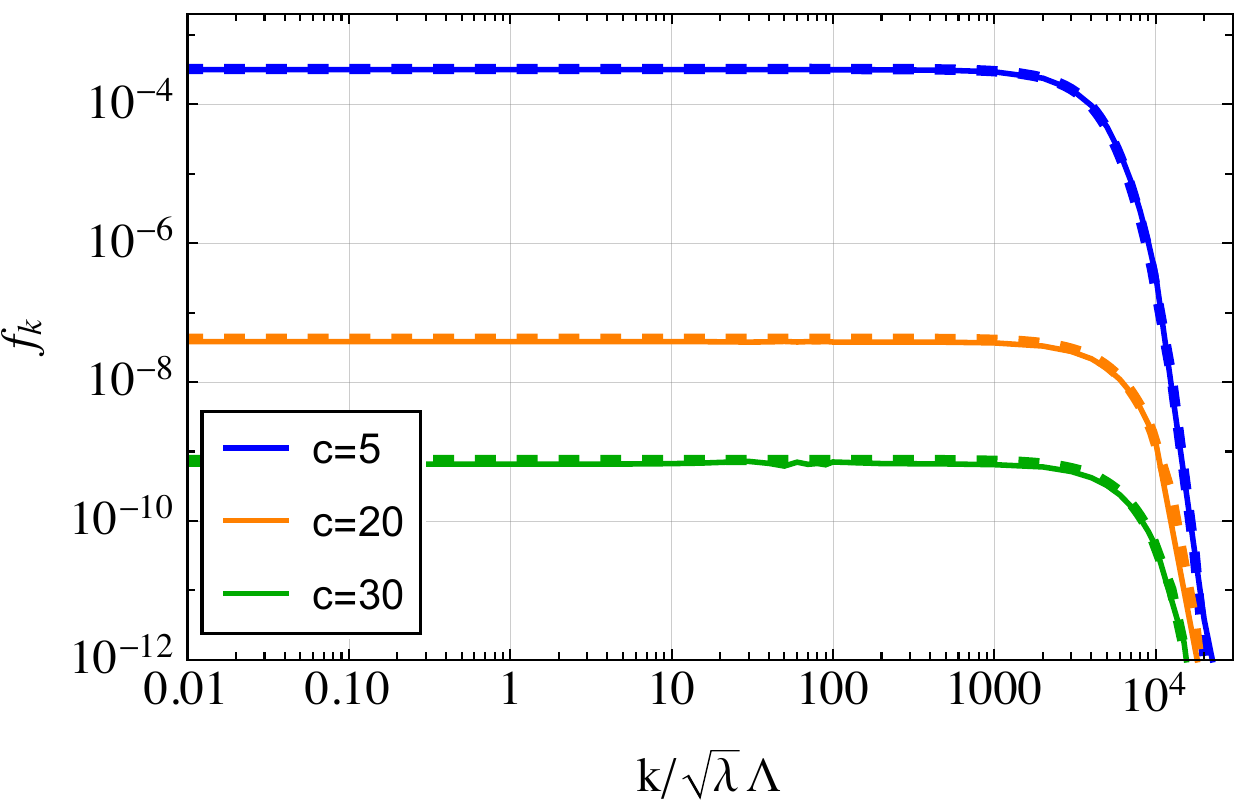}
	\includegraphics[width=0.45\linewidth]{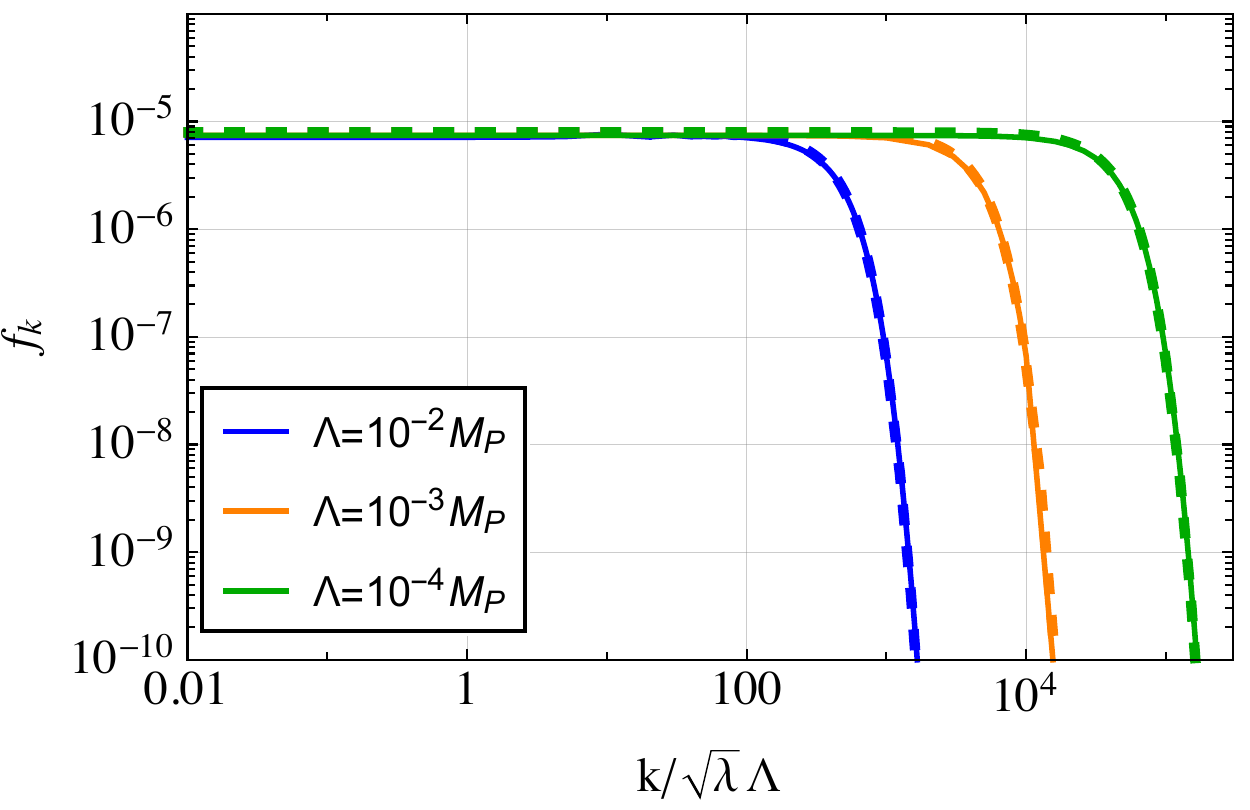}
	\caption{\small Particle number spectra for $\chi$-particles with a mass with a single pole. Solid lines correspond to the full numerical solution, dashed lines to the semiclassical approximation. The initial amplitude is $\Phi = 2\sqrt{M_P \Lambda}$. \\ \emph{Left:} $\Lambda = 10^{-3} M_P$, $\tilde c = 0.05 $ and $c=5,20,30$, (blue, orange, green, respectively). \\ \emph{Right:} $c = 10$, $\tilde c = 0.05 $ and $\Lambda = 10^{-2} M_P,10^{-3} M_P,10^{-4} M_P$, (blue, orange, green, respectively).}
	\label{fig:simple_pole}
\end{figure}

For comparison, we also solve Eqs.~\eqref{eq:eom_inflaton_t} and~\eqref{eq:alpha_beta_t} numerically, using a quartic potential
\begin{align*}
	V(\phi) = \frac{\lambda}{4} \phi^4\, .
\end{align*}
 We add an additional term to the mass of the $\chi$-particle
\begin{align}
	m_\chi^2 = c\frac{\dot \phi^2}{\Lambda^2} + 2\tilde c \frac{V(\phi)}{\Lambda^2}\, ,
\end{align}
which is irrelevant for the particle production around $\phi \sim 0$, but prevents adiabaticity violation as $\dot \phi \sim 0$, $k\rightarrow 0$. We have rescaled the momentum $k$ by $\sqrt\lambda \Lambda$, such that all dependence on $\lambda$ drops out. Here, and in the following, we evaluate the numerical spectrum at the point $\phi=\phi_*$, which is the first turning point of the inflaton, \emph{i.e.} the point $\dot \phi =0$ after the first zero crossing. In the case where we neglect expansion, $\phi_* = -\Phi$. In Fig.~\ref{fig:simple_pole}, we compare the numerical solution to the semiclassical formula~\eqref{eq:pia_simple}. As one can see, the formula~\eqref{eq:pia_simple} reproduces the numerical results extremely well.
This demonstrates the power of our semiclassical analysis. 

Eq.~\eqref{eq:pia_simple} indicates that the spectrum depends only on the height of the inflaton potential $V(\Phi)$ 
and not on its detailed structure.
We numerically computed the spectrum with a quadratic potential instead of quartic,
and confirmed that the spectrum is indeed intact as long as one takes the coefficient  
of the quadratic potential (or the inflaton mass)
such that $V(\Phi)$ is the same as the quartic case.

We point out, that as $c$ increases or $\tilde c$ decreases, the correspondence between the semiclassical approximation and the numerical result becomes worse at small $k$, as we will see explicitly in Sec.~\ref{sec:specmodels}. The cause is particle production by ordinary parametric resonance around $\dot\phi =0$, 
affecting the region of small $k$.\footnote{
	One often assumes an interaction of the form $\chi^2 \phi^2$,
	and then the parametric resonance happens 
	at $\phi = 0$ since the adiabaticity is broken at that point.
	In our case, since the interaction is of the form $\chi^2 \dot{\phi}^2$, the parametric resonance (if any)
	happens at $\dot{\phi} = 0$, not $\phi = 0$. 
	In other words, the phase is shifted by a quarter of
	the oscillation period.
} As this type of resonance is well studied in the literature, we will not pursue it any further. 

\subsubsection*{Example~2: mass with poles of order two}

Next, we take the effective mass term as
\begin{align}
	m_\chi^2 = c \frac{\dot{\phi}^2}{\Lambda^2 + \phi^2},
	\label{eq:double_pole_dphi}
\end{align}
where $c$ is again a numerical factor.
By using Eq.~\eqref{eq:rki_dphi_by_phi}, this is expressed as
\begin{align}
	m_\chi^2 \simeq 2c\frac{V(\Phi)}{\Lambda^2}\frac{1}{\left(1 + \phi^2/\Lambda^2\right)^2},
\end{align}
and hence it has a pair of poles of order two.
The quantities necessary for the semiclassical analysis are
\begin{align}
	\frac{1}{\omega_k^2}\frac{d\omega_k^2}{d\bar{\phi}}
	&= -\frac{4\bar{\phi}}{1+\bar{\phi}^2}\frac{1}{\bar{k}^2\left(1+\bar{\phi}^2\right)^2 + 1}, \\
	\frac{\omega_k^2}{\dot{\bar{\phi}}^2} &= c \left[\bar{k}^2\left(1+\bar{\phi}^2\right) + \frac{1}{1+\bar{\phi}^2}\right],
\end{align}
where we again defined the dimensionless parameters as
\begin{align}
	\bar{\phi} \equiv \frac{\phi}{\Lambda},
	\quad
	\bar{k}^2 \equiv \frac{1}{2c} \frac{k^2 \Lambda^2}{V(\Phi)}.
\end{align}
The prefactor of the integral has poles at
\begin{align}
	\bar{\phi} = \pm i, 
	\quad
	\pm i \sqrt{1\pm\frac{i}{\bar{k}}},
\end{align}
and hence Eq.~\eqref{eq:pia_general} reads
\begin{align}
	f_k \simeq \frac{\pi^2}{4}
	\abs{\exp\left[-2\sqrt{c}\int_0^{\bar{\phi}_+} d\bar{\phi}\, F \right]
	+ \exp\left[-2\sqrt{c}\int_0^{\bar{\phi}_-} d\bar{\phi}\, F \right]
	- 2\exp\left[-2\sqrt{c}\int_0^{1} d\bar{\phi}\, F \right]
	}^2,
	\label{eq:pia_order_two}
\end{align}
where we defined
\begin{align}
	F\left(\bar{k},\bar{\phi}\right) \equiv \sqrt{\bar{k}^2\left(1-\bar{\phi}^2\right) + \frac{1}{1-\bar{\phi}^2}},
	\quad
	\bar{\phi}_{\pm} = \sqrt{1\pm \frac{i}{\bar{k}}}.
\end{align}
\begin{figure}[t]
	\centering
 	\includegraphics[width=0.45\linewidth]{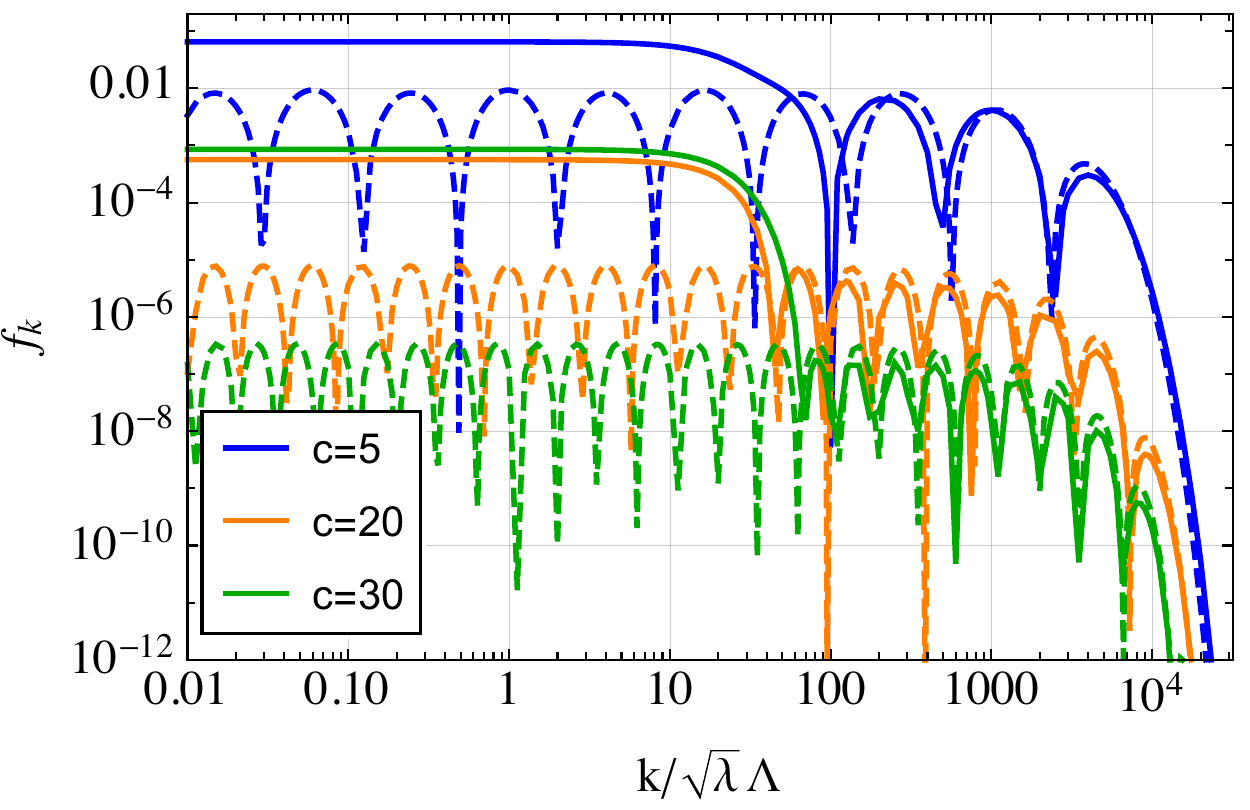}
	\includegraphics[width=0.45\linewidth]{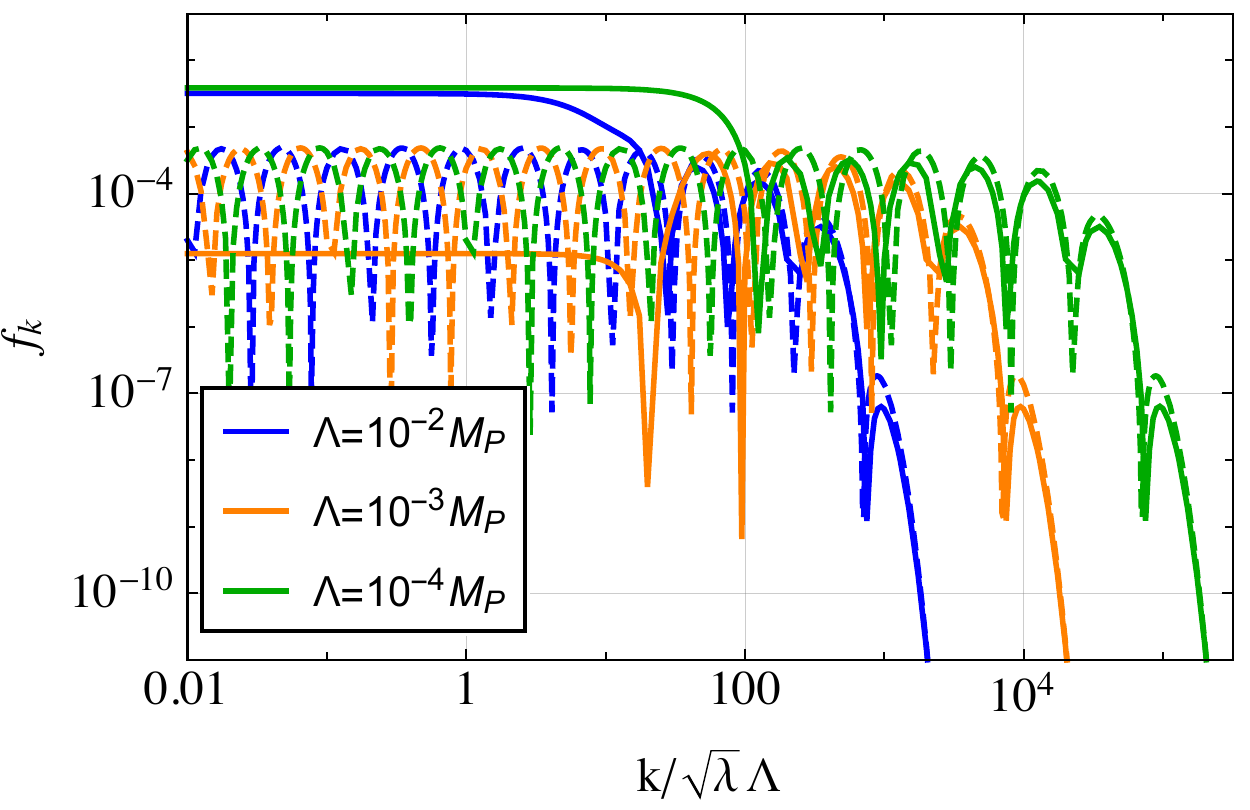}
	\caption{\small Particle number spectra for $\chi$-particles with a mass with a double pole. Solid lines correspond to the full numerical solution, dashed lines to the semiclassical approximation. The initial amplitude is $\Phi = 2\sqrt{M_P \Lambda}$ and the numerical spectra are evaluated at $\phi = \phi_*$. \\ \emph{Left:} $\Lambda = 10^{-3} M_P$, $\tilde c = 0.05 $ and $c=5,20,30$, (blue, orange, green, respectively). \\ \emph{Right:} $c = 10$, $\tilde c = 0.05 $ and $\Lambda = 10^{-2} M_P,10^{-3} M_P,10^{-4} M_P$, (blue, orange, green, respectively).}
	\label{fig:pole_of_order_two}
\end{figure}

In Fig.~\ref{fig:pole_of_order_two}, 
we compare the full numerical results 
with the semiclassical formula~\eqref{eq:pia_order_two} for a quartic potential. To temper the divergence at $\dot \phi \sim 0$, $k\rightarrow 0$ we again add a second term to the mass
\begin{align}
	m_\chi^2 = c\frac{\dot \phi^2}{\Lambda^2 + \phi^2} + 2\tilde c \frac{V(\phi)}{\Lambda^2}\, .
\end{align}
As one can see, the formula~\eqref{eq:pia_order_two}
reproduces the numerical results well, especially for large $k$,
including the oscillatory behavior. 
This oscillation originates from an interference between $\bar{\phi}_\pm$
which are generally complex. For small $k$ the ordinary parametric resonance around $\dot \phi =0$ causes additional particle production. In practice, this region is less important, as the contribution to the energy density of a mode with momentum $k$ scales as $\rho_k \propto k^3 \omega_k f_k$. The total energy density is thus dominated by the modes with large $k$, for which the semiclassical approximation works very well. 
Finally, we again checked that the large $k$ modes do not change if we use a quadratic inflaton potential instead of quartic
as long as $V(\Phi)$ is the same.

\section{Unitarity violation during preheating}
\label{sec:preheating_unitarity}
Recently, particle production after inflation, caused by the curvature of the scalar field target space, has been actively studied in the literature, in particular in the context of Higgs inflation.\footnote{
	See, \emph{e.g.}, Ref.~\cite{Renaux-Petel:2015mga} 
	for particle production from the target space curvature during inflation.
} Ref.~\cite{Ema:2016dny} found that the effective mass term of Goldstone modes exhibits a ``spiky" feature 
(see also Ref.~\cite{DeCross:2015uza}) in Higgs inflation, 
whose origin is now understood 
as the target space curvature in the Einstein frame~\cite{DeCross:2016cbs,Sfakianakis:2018lzf}.
The spike is of great phenomenological importance since it produces Goldstone modes 
whose energy scale can be greater than the cut-off scale of the theory, causing unitarity violation.\footnote{
	Longitudinal gauge bosons play the same role as the Goldstone modes in a gauged model
	due to the equivalence theorem.
} This indicates that one cannot determine the reheating temperature and hence the inflationary observables such as the spectral index and the tensor-to-scalar ratio within the validity of the theory.
Alternatively, one can put an upper bound on the nonminimal coupling $\xi$ between the Higgs and the Ricci scalar 
as $\xi \lesssim \mathcal{O}(100)$ to avoid the unitarity violation during preheating.

Since there are many inflation models other than Higgs inflation with curved target spaces, 
it is natural to ask if a similar phenomenon happens in those models.
In the following, we provide a general condition for 
a mass term from a curved target space to cause unitarity violation.
We will see that not only the target space curvature but also 
the height of the inflaton potential at the end of inflation plays an important role.
We then consider four inflation models as examples, \emph{i.e.}, 
running kinetic inflation~\cite{Takahashi:2010ky,Nakayama:2010kt}, 
Higgs inflation~\cite{Futamase:1987ua,CervantesCota:1995tz,Bezrukov:2007ep}, 
$\alpha$-attractor inflation~\cite{Kallosh:2013hoa,Kallosh:2013yoa} 
and Higgs-Palatini inflation~\cite{Bauer:2008zj,Bauer:2010jg},
and study particle production from the target space curvature.
We will see that the method we developed in Sec.~\ref{sec:pp_field_space} provides a useful estimation
of the spectrum of produced particles.
In this section, we include the Hubble expansion in our computation for completeness,
although it is of little importance for our purpose.

\subsection{Mass term from target space curvature}
\label{subsec:mass_curvature}

In this subsection, we review the covariant formalism and remind readers that the target space curvature induces a mass term~\cite{Sasaki:1995aw,GrootNibbelink:2000vx,GrootNibbelink:2001qt,Langlois:2008mn,Peterson:2010np,Gong:2011uw,Kaiser:2012ak}.\footnote{
	The target space in this formalism should be understood as defined only in the Einstein frame
	as it is variant under the frame transformation.
	An alternative frame-invariant definition of the target space is possible by including the conformal mode of
	the metric~\cite{Ema:2020zvg}.
} Let us begin with the following action in the Einstein frame:
\begin{align}
	S = \int d^4x \sqrt{-g} \left[
	\frac{M_P^2}{2}R 
	+ \frac{1}{2}h_{ab}\,g^{\mu\nu}\partial_\mu \phi^a \partial_\nu \phi^b
	- V
	\right],
\end{align}
where $M_P$ is the reduced Planck mass scale,
$g_{\mu\nu}$ is the spacetime metric with $g$ its determinant, 
$h_{ab} = h_{ab}(\phi)$ is the target space metric and $\phi^a$ is a scalar field with $a$ labeling its flavor.

Assuming the Friedmann-Lema{\^i}tre-Robertson-Walker (FLRW) metric,
the background equation of motion is given by
\begin{align}
	0 &= \frac{D \dot\phi_0^a}{Dt} + 3H\dot{\phi}^a_0 + h^{ab}V_b, \\
	H^2 &= \frac{1}{3M_P^2}\left[\frac{h_{ab}}{2}\dot{\phi}^a_0\dot{\phi}^b_0 + V\right], \\
	\dot{H} &= -\frac{1}{2M_P^2}h_{ab} \dot{\phi}^a_0 \dot{\phi}^b_0,
\end{align}
where the subscript ``0" is given to background quantities, $H$ is the Hubble parameter,
and $V_b$ is a short-hand notation for $\partial V/\partial \phi^b$.
The covariant derivative is defined as
\begin{align}
	\frac{D v^a}{Dt} &= \dot{v}^a + \Gamma^{a}_{bc} \dot{\phi}^b_0 v^c,
\end{align}
where $v^a$ is an arbitrary target space vector field and
$\Gamma^{a}_{bc}$ is the Christoffel symbol constructed from $h_{ab}$. 

We may fix the gauge degrees of freedom of the general coordinate transformation as
\begin{align}
	ds^2 = \mathcal{N}^2 dt^2 - a^2\left(dx^i + \beta^i dt\right)\left(dx^i + \beta^i dt\right),
\end{align}
where $\mathcal{N}$ is the lapse function, $\beta^i$ is the shift vector and $a$ is the scale factor.
Here we ignored the tensor part since it is irrelevant for our discussion.
The equations of motion of $\mathcal{N}$ and $\beta^i$ provide constraint equations, 
which are solved to leading order in perturbations as
\begin{align}
	\mathcal{N} - 1 &= \frac{H}{\dot{\phi}_0}\epsilon T_a \varphi^a,
\end{align}
where
\begin{align}
 	\varphi^a \equiv \phi^a - \phi^a_0,
 	\quad
	\dot{\phi}_0 = \sqrt{h_{ab}\dot{\phi}^a_0 \dot{\phi}^b_0},
	\quad 
	T^a = \frac{\dot{\phi}^a_0}{\dot{\phi}_0},
	\quad
	\epsilon = \frac{\dot{\phi}_0^2}{2M_P^2 H^2},
\end{align}
and the flavor indices are lowered by $h_{ab}$.
Here we do not show the solution of $\beta^i$ explicitly since
the quadratic action is linear in $\beta^i$ and hence it drops independently of its explicit form
after substituting the above form of $\mathcal{N}$.
The linearized equation of motion of the perturbation is then given by
\begin{align}
	0 = \frac{D^2 \varphi^a}{Dt^2} + 3H \frac{D\varphi^a}{Dt} 
	- \frac{1}{a^2}\partial_i^2 \varphi^a + h^{ab}M_{bc}^2\varphi^c,
\end{align}
where~\cite{Achucarro:2010da}
\begin{align}
	M_{ab}^2 &= \nabla_b V_a - \dot{\phi}^c_0 \dot{\phi}^d_0 R_{acdb} 
	+ \frac{2H}{\dot{\phi}_0}\epsilon \left(V_a T_b + V_b T_a\right)
	+ 2\left(3-\epsilon\right)\epsilon H^2 T_a T_b,
	\label{eq:covariant_mass_general} 
\end{align}
and all the geometrical quantities are constructed from $h_{ab}$.
Note in particular that the second term of Eq.~\eqref{eq:covariant_mass_general} originates from the target space curvature $R_{acdb}$.

From now, for simplicity, we consider the two-field case,
with the inflaton $\phi$ and another particle $\tilde{\chi}$.
We assume that $\tilde{\chi}$ does not have any background field value,
and focus on production of $\tilde{\chi}$. 
Thus, in the following, we omit the subscript ``$0$" as $\phi$ is always a background inflaton field 
and $\tilde{\chi}$ (or equivalently~$\chi$ defined below) is always a perturbation.
Assuming a $\mathbb{Z}_2$ symmetry under which only $\tilde{\chi}$ is odd, 
the mode equation of $\tilde{\chi}$ is always decoupled from the fluctuation of the inflaton.\footnote{
	This is because we assume that $\tilde{\chi}$ does not have any background field value 
	and hence the $\mathbb{Z}_2$ symmetry is not spontaneously broken.
} We may redefine $\tilde{\chi}$ as
\begin{align}
	\chi \equiv \sqrt{h_{\chi\chi}}\,\tilde{\chi},
\end{align}
where $h_{\chi\chi}$ here is a function of only the background field $\phi$.
Then the equation of motion simplifies significantly to
\begin{align}
	0 = \ddot{{\chi}} + 3H\dot{{\chi}} - \frac{1}{a^2}\partial_i^2 {\chi} 
	+ m_{\chi}^2\chi,
	\label{eq:chi_eom}
\end{align}
where
\begin{align}
	m_\chi^2 = \nabla^\chi V_\chi - \dot{\phi}^2 {R^{\chi}}_{\phi \phi \chi}.
	\label{eq:chi_mass}
\end{align}
Here we used the fact that\footnote{
	Note that this holds even if $\partial_\chi h_{\phi \chi} \neq 0$
	after substituting the background field configuration.
}
\begin{align}
	\dot{\chi} = \sqrt{h_{\chi\chi}}\frac{D\tilde{\chi}}{Dt},
\end{align}
and that the $\mathbb{Z}_2$ symmetry ensures that $V_\chi = 0$ at the background level.
The second term of Eq.~\eqref{eq:chi_mass} induces the spiky feature
and its geometrical nature is clear in this formalism.

It is then a standard exercise to move to Fourier space and quantize the modes.
The equations of motion of the Bogolibuv coefficients $\alpha_k$ and $\beta_k$ are given by
\begin{align}
	\frac{d{\alpha}_k}{d\tau} &= \frac{1}{4\omega_k^2}\frac{d \omega_k^2}{d\tau} \beta_k e^{2i\int^\tau d\tau\, \omega_k},
	\quad
	\frac{d{\beta}_k}{d\tau} = \frac{1}{4\omega_k^2}\frac{d \omega_k^2}{d\tau} \alpha_k e^{-2i\int^\tau d\tau\, \omega_k}.
	\label{eq:alpha_beta_tau}
\end{align}
where $\tau$ is comoving time.
The frequency is given by
\begin{align}
	\omega_k^2 = k^2 + a^2 m_\chi^2 - \frac{1}{a}\frac{d^2a}{d\tau^2},
\end{align}
with $k$ understood to be the comoving momentum.
The comoving occupation number is given by
\begin{align}
	f_k = \abs{\beta_k}^2.
	\label{eq:fchi_tau}
\end{align}

\subsection{Condition for unitarity violation}
\label{subsec:unitarity}

As we mentioned, it is known that the mass induced from the target space curvature 
causes unitarity violation during preheating in Higgs inflation.
On the other hand, although having a curved target space,
unitarity violation is not observed in, \emph{e.g.},
$\alpha$-attractor inflation and Higgs-Palatini inflation 
in an explicit analysis of these models~\cite{Krajewski:2018moi,Iarygina:2018kee,Iarygina:2020dwe,Rubio:2019ypq}.
This indicates that a nontrivial target space curvature alone is not sufficient
for unitarity violation. The purpose of this subsection is to derive a handy condition of when we expect unitarity violation
from the target space curvature during preheating.

In this work, we assume that unitarity is violated during preheating if the momentum of the mode which contributes most dominantly to the energy density, $k_\text{max}$, is larger than the unitarity cut-off scale.
Thus, we estimate $k_\mathrm{max}$ of particle production from the target space curvature 
in a model-independent way in the following.
Let $\Lambda$ denote the typical mass scale of the target space curvature around the origin.
Since scattering amplitudes of scalar fields depend on the target space curvature
(see, \emph{e.g.}, Ref.~\cite{Alonso:2015fsp} and references therein),
this scale $\Lambda$ corresponds to the (small field) cut-off scale of the theory at the same time;\footnote{
	Strictly speaking, gravitons also contribute to scattering amplitudes,
	and hence it is the target space curvature that contains not only scalar fields 
	but also the (scalar part of) gravity defined in Ref.~\cite{Ema:2020zvg} that determines the cut-off scale.
	Practically, however, this point is not important for $\Lambda \ll M_P$
	as long as one works solely in the Einstein frame as we do in this paper. 
}
we will comment on this choice at the end of this subsection. We expect that the dynamics of the inflaton is drastically different 
in the regimes $\abs{\phi} \gtrsim \Lambda$ and $\abs{\phi} \lesssim \Lambda$.
The duration of the inflaton passing the regime $\abs{\phi} \lesssim \Lambda$, $\Delta t$, is estimated as
\begin{align}
	\left.\dot{\phi}\,\Delta t\right\rvert_{\abs{\phi}\lesssim \Lambda} \sim \Lambda,
\end{align}
and hence we expect the typical energy scale of particles produced by this change as
\begin{align}
	\left(\frac{k}{a_*}\right)_\mathrm{max} \sim \Delta t^{-1} \sim \left.\frac{\dot{\phi}}{\Lambda}\right\rvert_{\abs{\phi}\lesssim \Lambda},
\end{align}
where $k$ is understood to be comoving 
and $a_*$ is the scale factor at the first turning point of the inflaton, at which $\phi = \phi_*$.
We further take the scalar kinetic terms canonical around the origin, 
$h_{\phi\phi} \simeq 1$ for $\abs{\phi} \lesssim \Lambda$.
Energy conservation then tells us that the inflaton velocity around the origin 
is related to the inflaton potential at the end of inflation via
\begin{align}
	\left. \dot{\phi}^2 \right\rvert_{\abs{\phi}\lesssim \Lambda} \sim V(\Phi),
\end{align}
where $\Phi$ is the inflaton amplitude at the end of inflation.
The condition for unitarity violation, $(k/a_*)_\mathrm{max} \gtrsim \Lambda$,
now reads
\begin{align}
	V(\Phi) \gtrsim \Lambda^4.
	\label{eq:unitarity_condition}
\end{align}
Thus, we expect that unitarity violation happens if the potential at the end of inflation is larger than the target space curvature around the origin.

We can derive the same condition for the examples in Sec.~\ref{subsec:pia_examples}.
There the momentum $k$ always appears in the combination
\begin{align}
	\bar{k}^2 = \frac{1}{2c} \frac{k^2 \Lambda^2}{V(\Phi)},
\end{align}
and hence the typical energy scale of the particle production is
\begin{align}
	\left(\frac{k}{a_*}\right)_\mathrm{max}^2 \sim \frac{V(\Phi)}{\Lambda^2},
\end{align}
where we assumed that $c$ is of order unity,
and reinterpret $k$ in Sec.~\ref{subsec:pia_examples} as $k/a_*$.
Requiring that $(k/a_*)_\mathrm{max}$ is larger than $\Lambda$, 
we arrive at the same condition.

Let us now comment on our choice of unitarity cut-off scale. Depending on the model, the cut-off scale can depend on the background field value, as is the case for Higgs inflation~\cite{Ferrara:2010in, Bezrukov:2010jz}. Although we evaluate the spectrum at the end point of the oscillation, particle number for the UV-modes is already well-defined at $|\phi| \sim \Lambda$. For this reason, we adopt the small-field value of the cut-off scale when we determine whether unitarity is violated.

In the following, we consider
running kinetic inflation, Higgs inflation, $\alpha$-attractor inflation and Higgs-Palatini inflation,
and study particle production and unitarity violation caused by the target space curvature.
We will see that Eq.~\eqref{eq:unitarity_condition} is satisfied for the former two models,
while it is violated for the latter two models,
explaining the different unitarity structures of these models during preheating.

\subsection{Analysis of specific inflation models}
\label{sec:specmodels}

From now, we consider
running kinetic inflation, Higgs inflation, $\alpha$-attractor inflation and Higgs-Palatini inflation in turn. 
We study the particle production and the unitarity structure of these models after one inflaton oscillation. 
It will turn out that the method we developed in Sec.~\ref{sec:pp_field_space} 
is a powerful tool to study the spectrum analytically. In contrast to Sec.~\ref{sec:pp_field_space}, we include the expansion of the universe in our computations for completeness, although it is unimportant for our main discussion.

\subsubsection*{Running kinetic inflation}

First, we consider running kinetic inflation.
We consider the following action
\begin{align}
	S = \int d^4 x \sqrt{-g} \left [ \frac{M_P^2}{2} R 
	+ \frac{1}{2}\left(1 + \frac{\phi^2}{\Lambda^2} - c_K \frac{\chi^2}{\Lambda^2}
	\right) \left(\partial \phi \right)^2
	+ \frac{1}{2}\left(\partial \chi \right)^2 
	- \left(1+c_V\frac{\chi^2}{\Lambda^2}\right)\frac{\lambda \phi^4}{4} \right],
	\label{eq:action_rki}
\end{align}
where we put couplings between $\phi$ and $\chi$ just by hand,
and assume that $c_K, c_V > 0$, since otherwise $\chi$ can be tachyonic.
If we instead identify the inflaton as, \emph{e.g.}, the Higgs
and consider production of the Goldstone modes,
the analysis would be closer to Higgs inflation which we discuss hereafter.
This model reduces to a chaotic inflation model with a quadratic potential in the large field region,
which is now disfavored by the CMB observation.
This can be overcome however by taking the exponent of $\phi$ in the prefactor of the kinetic term larger. 
Although our analysis, in particular the argument in Sec.~\ref{sec:pp_field_space}, equally applies to such a case, for simplicity
we will stick to Eq.~\eqref{eq:action_rki} in the following.

Obviously this model has the target space metric
\begin{align}
	h_{ab} &= \mathrm{diag}\left(1 +  \frac{\phi^2}{\Lambda^2} - c_K \frac{\chi^2}{\Lambda^2}, 1\right).
\end{align}
The background equations of motion are given by
\begin{align}
	0 &= \left(1+\frac{\phi^2}{\Lambda^2}\right)\ddot{\phi} + 3 H \left(1+\frac{\phi^2}{\Lambda^2}\right) \dot{\phi}
	+ \frac{\phi}{\Lambda^2}\dot{\phi}^2 + \lambda \phi^3, \\
	H^2 &= \frac{1}{3M_P^2}\left[\frac{1}{2}\left(1+\frac{\phi^2}{\Lambda^2}\right)\dot{\phi}^2 
	+ \frac{\lambda}{4}\phi^4\right], \\
	\dot{H} &= -\frac{1}{2M_P^2}\left(1+\frac{\phi^2}{\Lambda^2}\right)\dot{\phi}^2\, .
\end{align}
Inflation happens for $\phi^2 \gtrsim M_P \Lambda$ in this model,
and the CMB normalization requires that
\begin{align}
	\lambda \Lambda^2 \sim 10^{-10} M_P^2,
	\label{eq:rki_CMB}
\end{align}
which indicates that $\Lambda/M_P \ll 1$ unless $\lambda$ is tiny.

We now apply the condition~\eqref{eq:unitarity_condition} to this model.
The height of the inflaton potential at the end of inflation is given by
\begin{align}
	V(\Phi) \sim \lambda M_P^2 \Lambda^2.
\end{align}
Since the cut-off scale of this theory is of order $\Lambda$ if we assume $c_K$ to be of order unity,
we obtain
\begin{align}
	\frac{V(\Phi)}{\Lambda^4} \sim \frac{\lambda M_P^2}{\Lambda^2}
	\sim 10^{-10}\left(\frac{M_P}{\Lambda}\right)^4,
	\label{eq:unitarity_rki}
\end{align}
where we used Eq.~\eqref{eq:rki_CMB} in the second similarity.
The ratio can be larger than unity for small $\Lambda$,
and thus we expect unitarity violation in this model.
To our knowledge, we are pointing out the possibility of unitarity violation during preheating
in running kinetic inflation here for the first time (except for a brief comment in Ref.~\cite{Ema:2016dny}).

\begin{figure}[t]
	\centering
 	\includegraphics[width=0.495\linewidth]{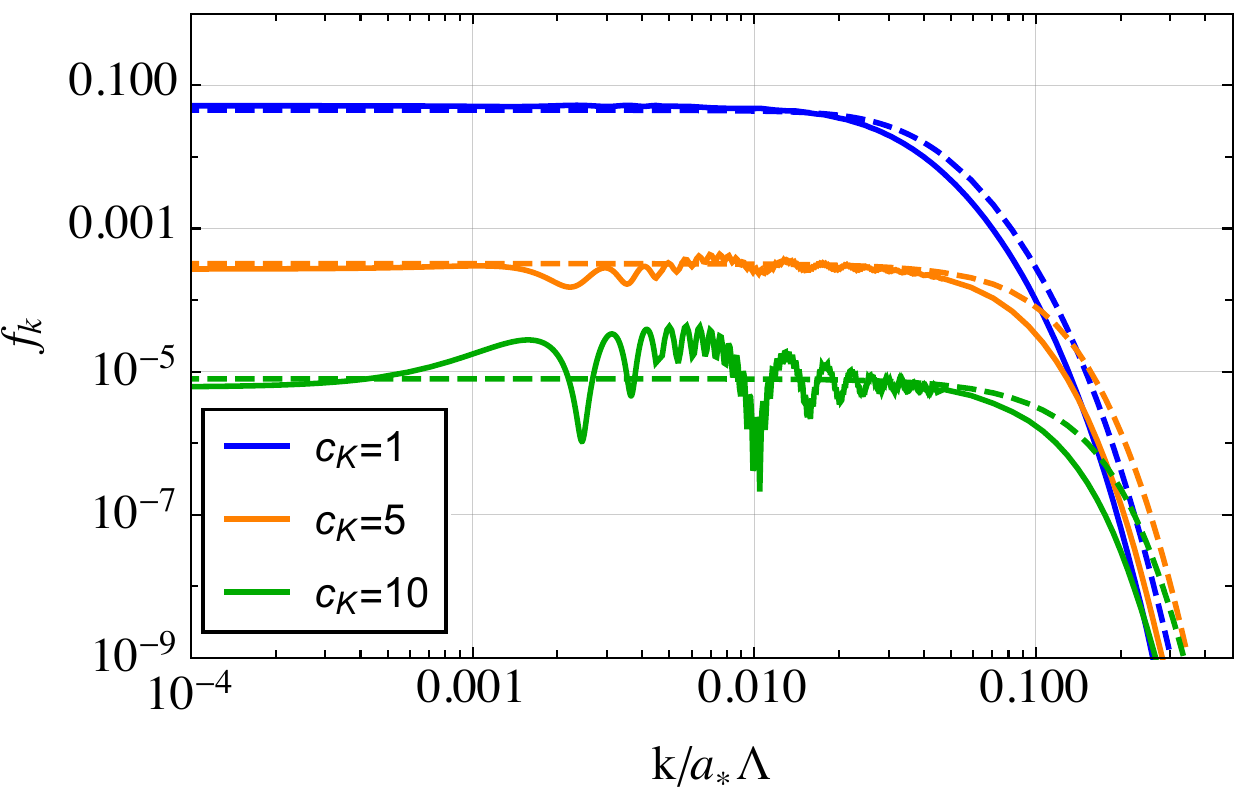}
	 \includegraphics[width=0.495\linewidth]{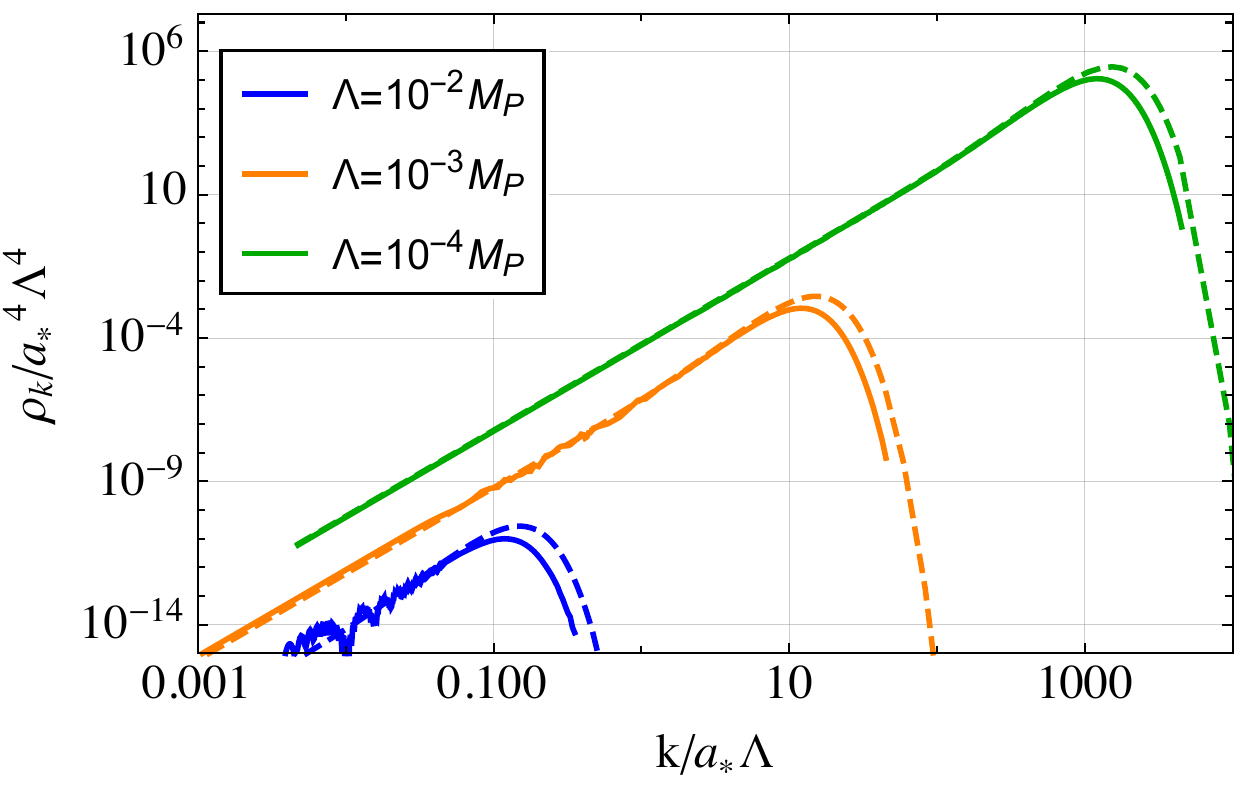}
	\caption{\small Particle number density and energy density for running kinetic inflation.
	The solid lines indicate the numerical solution to the mode equation, the dashed line the semiclassical approximation. \\ \emph{Left:} particle number density for $\Lambda = 10^{-2} M_P$, $c_V = 0.1$ and $c_K = 1, 5, 10$, (blue, orange, green respectively).\\ \emph{Right:} energy density for $c_K = 10$, $c_V=0.1$ and $\Lambda = 10^{-2}M_P, 10^{-3}M_P, 10^{-4}M_P$, (blue, orange, green respectively).}
	\label{fig:pp_rki}
\end{figure}

We now study the spectrum of the produced particles in detail 
with the method developed in Sec.~\ref{sec:pp_field_space}.
The relevant geometrical quantities are given by
\begin{align}
	{R^\chi}_{\phi\phi\chi} &= -\frac{c_K}{\Lambda^2},
	\quad
	\nabla^\chi V_\chi = c_V \frac{\lambda \phi^4}{4\Lambda^2},
\end{align}
and hence the effective mass is given by
\begin{align}
	m_\chi^2 &= \frac{1}{\Lambda^2}\left[c_K\dot{\phi}^2 + c_V\frac{\lambda \phi^4}{2}\right].
	\label{eq:chi_mass_rki}
\end{align}
Numerically we can simply substitute this expression and solve Eq.~\eqref{eq:alpha_beta_tau}
with the background equations of motion. The spectrum is then given by Eq.~\eqref{eq:fchi_tau}.

In order to interpret the numerical results analytically,
we ignore $c_V$ and the Hubble parameter
since they are expected to be subdominant for the particle production.
This simplifies the mass as
\begin{align}
	m_\chi^2 \simeq c_K\frac{\dot{\phi}^2}{\Lambda^2}.
\end{align}
If we ignore the Hubble expansion, energy conservation tells us that
\begin{align}
	\dot{\phi}^2
	\simeq
	\frac{\lambda}{2}\frac{\Phi^4}{1 + \phi^2/\Lambda^2},
\end{align}
where $\Phi = \mathcal{O}(\sqrt{M_P \Lambda})$ is the inflaton amplitude at the end of inflation.
These relations correspond to the first example in Sec.~\ref{subsec:pia_examples},
and hence we can directly apply the analysis there to this model.
As a result, we estimate the occupation number as
\begin{align}
	f_k \simeq 
	\frac{\pi^4}{4}\abs{
	\exp\left[-\sqrt{c_K}\left(1+\left(\bar{k}+\frac{1}{\bar{k}}\right)\arctan\left(\bar{k}\right)\right)\right]
	- \exp\left[-\frac{\pi \sqrt{c_K}}{2}\frac{\bar{k}^2+1}{\bar{k}}\right]
	}^2.
	\label{eq:pia_rki}
\end{align}
where 
\begin{align}
	\bar{\phi} = \frac{\phi}{\Lambda},
	\quad
	\bar{k}^2 \equiv \frac{2}{c_K} \frac{k^2 \Lambda^2/a_*^2}{\lambda \Phi^4}.
\end{align}
In our semiclassical formula, 
we always take $\Phi$ as the inflaton field value at the point $\epsilon \equiv -\dot{H}/H^2 =1$.
Note also that we use the scale factor at the first turning point $a_*$ in the definition of $\bar{k}$.
Since the argument in Sec.~\ref{sec:pp_field_space} does not include the cosmic expansion,
we have small uncertainties related to the choice of $\Phi$ and $a$ here,
which we cannot resolve within our semiclassical analysis without including the Hubble expansion.

In Fig.~\ref{fig:pp_rki}, we compare the numerical results with the analytical estimate~\eqref{eq:pia_rki}, choosing
\begin{align}
	\lambda = 10^{-10}\left( \frac{M_P}{\Lambda} \right)^2\, .
\end{align}
Note that we no longer rescale the physical momentum by $\sqrt{\lambda}\Lambda$, but just by the unitarity cut-off scale $\Lambda$, such that it is easy to see if unitarity is violated. The result now depends on what value is chosen for $\lambda$. We initialize the background equation sufficiently early, such that it follows the attractor solution towards the end of inflation (set by the slow roll parameter $\epsilon =1$). We initialize the mode functions 2 $e$-folds before the end of inflation. At this point, all modes are adiabatic, \emph{i.e.} $|\dot \omega_k/\omega_k^2| \ll 1$, and particle number is well-defined. The particle number and energy density shown in Fig.~\ref{fig:pp_rki} are evaluated at $\phi = \phi_*$, when all modes are adiabatic again.\footnote{Note that we can not choose $c_V$ too small, as otherwise adiabaticity is violated when $\dot \phi \sim 0$ for the IR modes. This is not the case for our chosen value of $c_V$.}

The left panel of Fig.~\ref{fig:pp_rki} shows that the semiclassical result approximates the numerics reasonably well, although the correspondence is somewhat worse than we saw in Fig.~\ref{fig:simple_pole} for the single pole case. Two factors cause the deviation. First, we expect some particle production from ordinary parametric resonance, which is not described by equation~\eqref{eq:pia_rki}. This contribution increases with $c_K$ and would also have occurred in the single pole case of Fig.~\ref{fig:simple_pole} if we had chosen a larger value of  $c$. Second, the inflaton amplitude decreases due to Hubble expansion, and the conserved quantity as defined in Eq.~\eqref{eq:conserved_quantity_general} is only approximately conserved.

\begin{figure}[t]
	\centering
 	\includegraphics[width=0.495\linewidth]{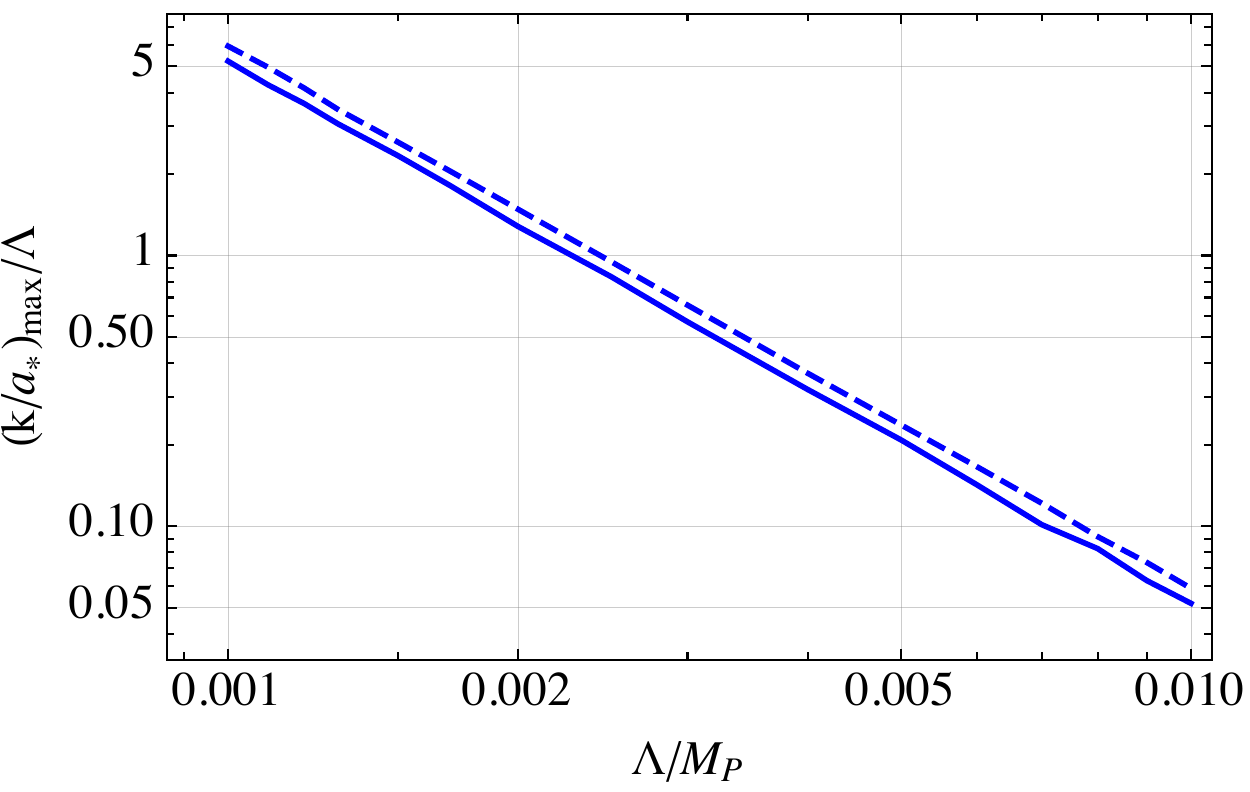}
	 \includegraphics[width=0.495\linewidth]{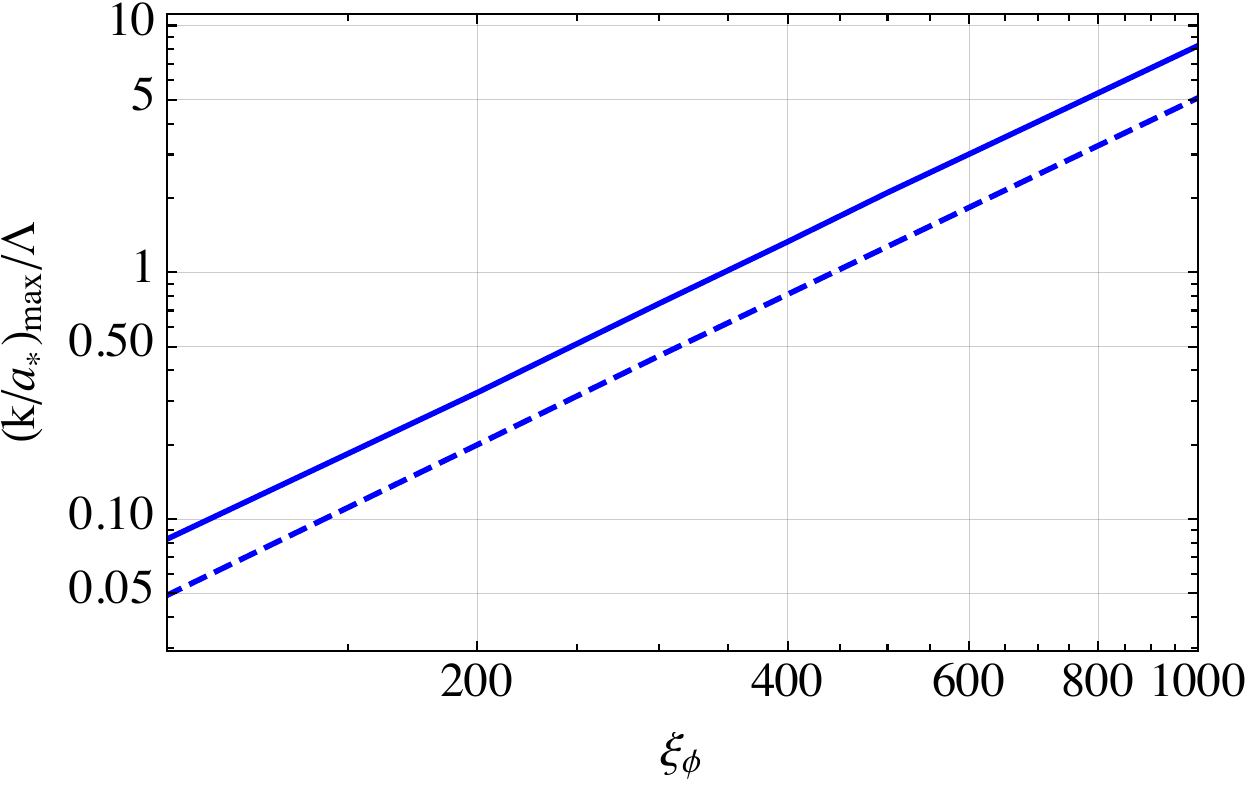}
	\caption{\small Ratio of $(k/a_*)_\text{max}$ to the unitarity cut-off scale for running kinetic inflation, with $c_K = 1$ and $c_V = 0.1$, (left) and Higgs inflation, with $\xi_\phi = \xi_\chi$, (right) (with $\Lambda = M_P/\xi_\phi$). The solid line shows the numerical result and the dashed line the semiclassical approximation.}
	\label{fig:kmax}
\end{figure}

Nevertheless, the right panel of Fig.~\ref{fig:pp_rki} shows that the semiclassical result gives an excellent estimate of $k_\text{max}$ and can thus be used to determine unitarity violation.
Here the vertical axis is the physical energy density $\rho_k/a_*^4$ per $\log k/a_*$, where $\rho_k \equiv k^3 \omega_k f_k/2\pi^2$ is the comoving energy density.
In the left panel of Fig.~\ref{fig:kmax}, we show the numerical results for the ratio of the maximum physical momentum $(k/a_*)_\text{max}$ and the cut-off scale. Assuming that unitarity is violated when this ratio is larger than 1, we find that unitarity is violated for $\Lambda \lesssim 2 \times 10^{-3} M_P$, consistent with our estimate of Eq.~\eqref{eq:unitarity_rki}.
We also note that the semiclassical formula works well even for $\Lambda = 10^{-2}M_P$,
a regime in which unitarity is not violated. The semiclassical formula is thus not only useful for the determination of the minimum value of $\Lambda$ to avoid unitarity violation, 
but also to study the spectrum in the case without unitarity violation.

Now let us make a brief comment on possible UV completions of the present model. If unitarity is violated during the preheating stage as seen above, it is necessary to consider a UV-completed theory to discuss preheating. 
One example of the renormalizable action is something as follows:
\begin{align}
	S=\int d^4x\sqrt{-g}\left[
	\frac{M_P^2}{2}R + \frac{1}{2}(\partial\psi)^2+ \frac{1}{2}(\partial\phi)^2+\frac{1}{2}(\partial\chi)^2- \frac{1}{2}(m \psi-c_\phi \phi^2-c_\chi\chi^2)^2 - V(\phi)\right],
	\label{RKI_UV1}
\end{align}
where $\psi$, $\phi$ and $\chi$ are real scalar fields, $c_\phi$ and $c_\chi$ are constants assumed to be positive.
We assume that $m$ is larger than any other scale appearing in $V(\phi)$. We may then integrate out $\psi$ by using the relation $m \psi = c_\phi\phi^2+c_\chi\chi^2$ and the inflationary trajectory is taken to be $m\psi= c_\phi\phi^2$ with $\chi=0$. Then the action reduces to the form of Eq.~(\ref{eq:action_rki}) with $\Lambda= m/(2c_\phi)$ and $c_K=c_\chi/(4c_\phi)$. 
Another example is
\begin{align}
	S=\int d^4x\sqrt{-g}\left[\frac{M_P^2}{2}R +
	\frac{1}{2}(\partial\psi)^2+ \frac{1}{2}(\partial\phi)^2+ \frac{1}{2}(\partial\chi)^2 - \frac{1}{4}(m^2-c_\psi\psi^2+c_\phi\phi^2+c_\chi\chi^2)^2 - V(\phi)\right],
	\label{RKI_UV2}
\end{align}
where we assume that $c_\phi$ and $c_\psi$ are positive and satisfy $c_\phi\gg c_\psi$ and again $m$ is larger than any other scale appearing in $V(\phi)$. Then, for $|\phi|\lesssim m/\sqrt{c_\phi}$, it reduces to the effective action (\ref{eq:action_rki}) with $\Lambda= \sqrt{c_\psi} m /c_\phi$ and $c_K=c_\chi/c_\phi$ after eliminating $\psi$ with the use of the constraint $c_\psi\psi^2=m^2+c_\phi\phi^2+c_\chi\chi^2$.
Both models give the same effective theory (\ref{eq:action_rki}) in a certain parameter range and field space, but phenomenological implications are quite different. In the model (\ref{RKI_UV1}), one should take into account the production of the heavy mode that would have been integrated out in the effective theory, but it may not have significant phenomenological impact as far as it decays into lighter degrees of freedom or the Standard Model sector quickly. 
On the other hand, in the model (\ref{RKI_UV2}), there is a possibility of domain wall formation if $V(\phi_{\rm ini}) \gtrsim m^4$ due to the multi-field scalar dynamics, since the action has a $\mathbb{Z}_2$ symmetry under which $\phi$ and $\psi$ change their sign and the final vacuum expectation values $(\phi,\psi)=(0,\sqrt{c_\psi} m)$ spontaneously break the $\mathbb{Z}_2$ symmetry. The formation of domain walls during the preheating is a cosmological disaster unless there are some mechanisms to make the domain walls unstable.
This inspection shows that one should be careful about the unitarity violation in the effective theory. Even if one can safely use the effective theory like (\ref{eq:action_rki}) during slow-roll inflation, the validity of that theory may not always be guaranteed for the analysis of (p)reheating. Once unitarity violation is observed, one should go back to the UV-completed theory, and the phenomenological consequences are sensitive to the concrete model of UV completion.

\subsubsection*{Higgs inflation}

Next we consider Higgs inflation.
We start from the Jordan frame action 
\begin{align}
	S = \int d^4 x \sqrt{-g_J} \left [ \frac{M_P^2}{2}\Omega^2 R_J 
	+ \frac{1}{2} g_J^{\mu\nu} \partial_\mu \phi^a \partial_\nu \phi^a 
	- V_J(\phi, \chi) \right]\, ,
	\label{eq:action_Higgs_inf}
\end{align}
where the subscripts ``$J$" denote quantities in the Jordan frame, $\phi^a = \phi, \chi$, and we define
\begin{align}
	\Omega^2(\phi,\chi) = 1 + \frac{\xi_\phi \phi^2 + \xi_\chi \chi^2}{M_P^2},
\end{align}
and
\begin{align}
	V_J(\phi,\chi) = \frac{\lambda_\phi}{4} \phi^4 + \frac{g^2}{2}\phi^2 \chi^2 + \frac{\lambda_\chi}{4}\chi^4\, .
\end{align}
Here we have slightly generalized the model, and the case $\xi_\chi = \xi_\phi$ and $\lambda_\chi = g^2 = \lambda_\phi$
corresponds to the original (abelian) Higgs inflation model.\footnote{
	The discussion below is the same for an SU(2) case
	as long as a linearized mode equation is concerned.
} We always assume that $\xi_\phi$ and $\xi_\chi$ are of comparable order.

It is convenient to go to the Einstein frame by rescaling the metric as 
\begin{align}
	g_{J\mu\nu} \rightarrow g_{\mu\nu} = \Omega^2 g_{J\mu\nu},
\end{align}
resulting in
\begin{align}
	S = \int d^4 x \sqrt{-g} \left[\frac{M_P^2}{2} R + \frac{1}{2} h_{ab} g^{\mu\nu} \partial_\mu \phi^a \partial_\mu \phi^b 
	-V(\phi, \chi) \right] \, ,
\end{align}
with the potential in the Einstein frame given by
\begin{align}
	V(\phi, \chi) = \frac{V_J(\phi, \chi)}{\Omega^4}\, ,
\end{align}
and the target space metric
\begin{align}
	h_{ab} &= \frac{1}{\Omega^4} 
	\left[\Omega^2 \delta_{ab} + \frac{3}{2}\partial_a \Omega^2 \partial_b \Omega^2\right] 
	= \frac{1}{\Omega^4} 
	\begin{pmatrix}
	\Omega^2 + \frac{6\xi_\phi^2 \phi^2}{M_P^2} & \frac{6 \xi_\phi \xi_\chi \phi \chi}{M_P^2} \\
	\frac{6 \xi_\phi \xi_\chi \phi \chi}{M_P^2} & \Omega^2 + \frac{6\xi_\chi^2 \chi^2}{M_P^2} 
	\end{pmatrix}\, .
\end{align}
For generic initial conditions, this model does not display strong turning, so without loss of generality, 
we can take $\phi$ to be the inflaton. 
The background value of $\chi$ will remain zero 
(this was checked on the lattice for $1<\xi_\phi<100$ \cite{Nguyen:2019kbm, vandeVis:2020qcp})
as long as the parameters satisfy
\begin{align}
	\frac{g^2}{\lambda_\phi} \geq \frac{\xi_\chi}{\xi_\phi},
\end{align}
which we assume to be the case (see Eq.~\eqref{eq:mass_pot_HI}).
This allows us to apply the argument of Sec.~\ref{subsec:mass_curvature} to this model.

In the single-field limit, the background equation of the inflaton field is
\begin{align}
	0 &= \ddot{\phi} + 3H \dot{\phi} 
	+ \left(\frac{\xi_\phi\left(1+6\xi_\phi\right)\phi}{1+\xi_\phi\left(1+6\xi_\phi\right)\phi^2/M_P^2}
	- \frac{2\xi_\phi \phi}{1 + \xi_\phi \phi^2/M_P^2}
	\right)\frac{\dot{\phi}^2}{M_P^2} \nonumber \\
	&+ \frac{\lambda_\phi \phi^3}{\left(1+\xi_\phi \phi^2/M_P^2\right)
	\left(1 + \xi_\phi\left(1+6\xi_\phi\right)\phi^2/M_P^2\right)},
\end{align}
and the Friedman equations are
\begin{align}
	H^2 = & \frac{1}{3M_P^2} 
	\left(\frac{1 + \xi_\phi \left(1+6\xi_\phi\right)\phi^2/M_P^2}{2(1 + \xi_\phi \phi^2/M_P^2)^2} \dot\phi^2 
	+ \frac{\lambda_\phi \phi^4}{4(1 + \xi_\phi \phi^2/M_P^2)^2} \right)\, , \\
	\dot H = & -\frac{ (1+ \xi_\phi\left(1+6\xi_\phi\right) \phi^2/M_P^2)}{2(1 + \xi_\phi \phi^2/M_P^2)^2}
	\frac{\dot \phi^2}{M_P^2}\, .
\end{align}
Higgs inflation is well known to be consistent with the CMB observation~\cite{Akrami:2018odb},
and the CMB normalization requires that
\begin{align}
	\frac{\xi_\phi^2}{\lambda_\phi} \simeq 2\times 10^9.
	\label{eq:Higgs_inf_CMB}
\end{align}
This indicates that $\xi_\phi \gg 1$ unless $\lambda_\phi$ is tiny at the inflationary scale.
This large value of $\xi_\phi$ is an active subject of discussion.
A notable consequence is that the cut-off scale of the theory becomes 
as low as $\Lambda \sim M_P/\xi_\phi$~\cite{Burgess:2009ea,Barbon:2009ya,Burgess:2010zq,Hertzberg:2010dc}.\footnote{
	As we mentioned above, we assume that $\xi_\chi$ and $\xi_\phi$ are comparable
	in this paper.
	The cut-off scales as $M_P/\sqrt{\xi_\chi \xi_\phi}$ for $\xi_\phi, \xi_\chi \gg 1$
	when $\xi_\phi$ and $\xi_\chi$ are not comparable,
	as can be seen in Eq.~\eqref{eq:mass_curvature_HI}.
	This is related to the fact that $\xi_\phi$ is physical only when $\chi$ is present 
	(without the inflaton potential)~\cite{Hertzberg:2010dc}.
} This is called the unitarity issue of Higgs inflation, and different viewpoints on this issue are discussed at the end of this subsection. Our point in this paper is that, even if the analysis during inflation is not spoiled due to the field-dependence of the cut-off~\cite{Ferrara:2010in, Bezrukov:2010jz},
unitarity can still be violated after inflation~\cite{Ema:2016dny}.
The preheating dynamics of Higgs inflation was studied in great detail in Refs.~\cite{Sfakianakis:2018lzf,Nguyen:2019kbm,vandeVis:2020qcp}.

Now we apply the condition~\eqref{eq:unitarity_condition} to this model.
Inflation happens for $\phi \gtrsim M_P/\sqrt{\xi_\phi}$,
and hence the height of the inflaton potential at the end of inflation is given by
\begin{align}
	V(\Phi) \sim \frac{\lambda_\phi M_P^4}{\xi_\phi^2}.
\end{align}
Since the cut-off scale of this theory is $\Lambda \sim M_P/\xi_\phi$, we obtain
\begin{align}
	\frac{V(\Phi)}{\Lambda^4} \sim \lambda_\phi \xi_\phi^2
	\sim 10^{-9}\xi_\phi^4,
	\label{eq:unitarity_Higgs_inf}
\end{align}
where we used Eq.~\eqref{eq:Higgs_inf_CMB} in the second similarity.
The ratio can be larger than unity for large $\xi_\phi$,
and thus the condition~\eqref{eq:unitarity_condition} indeed
signals unitarity violation in this model.

\begin{figure}[t]
	\centering
 	\includegraphics[width=0.495\linewidth]{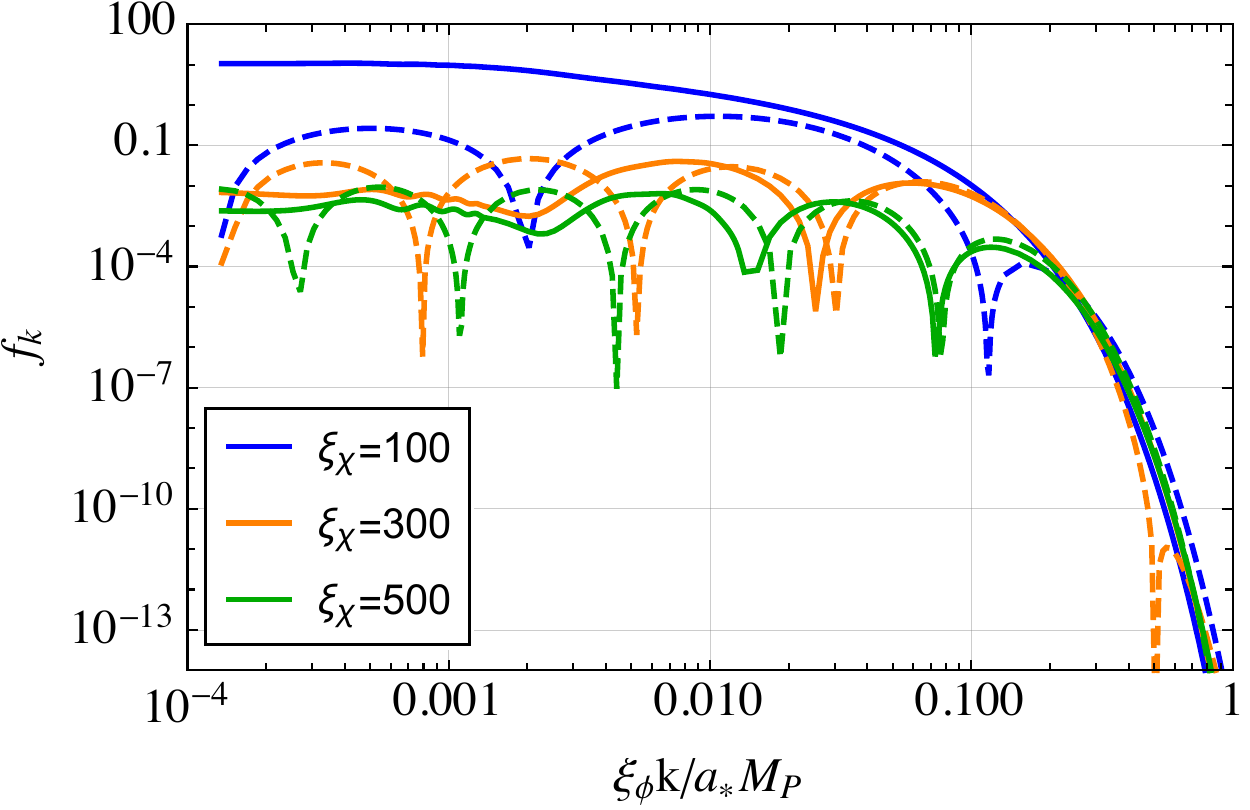}
	\includegraphics[width=0.495\linewidth]{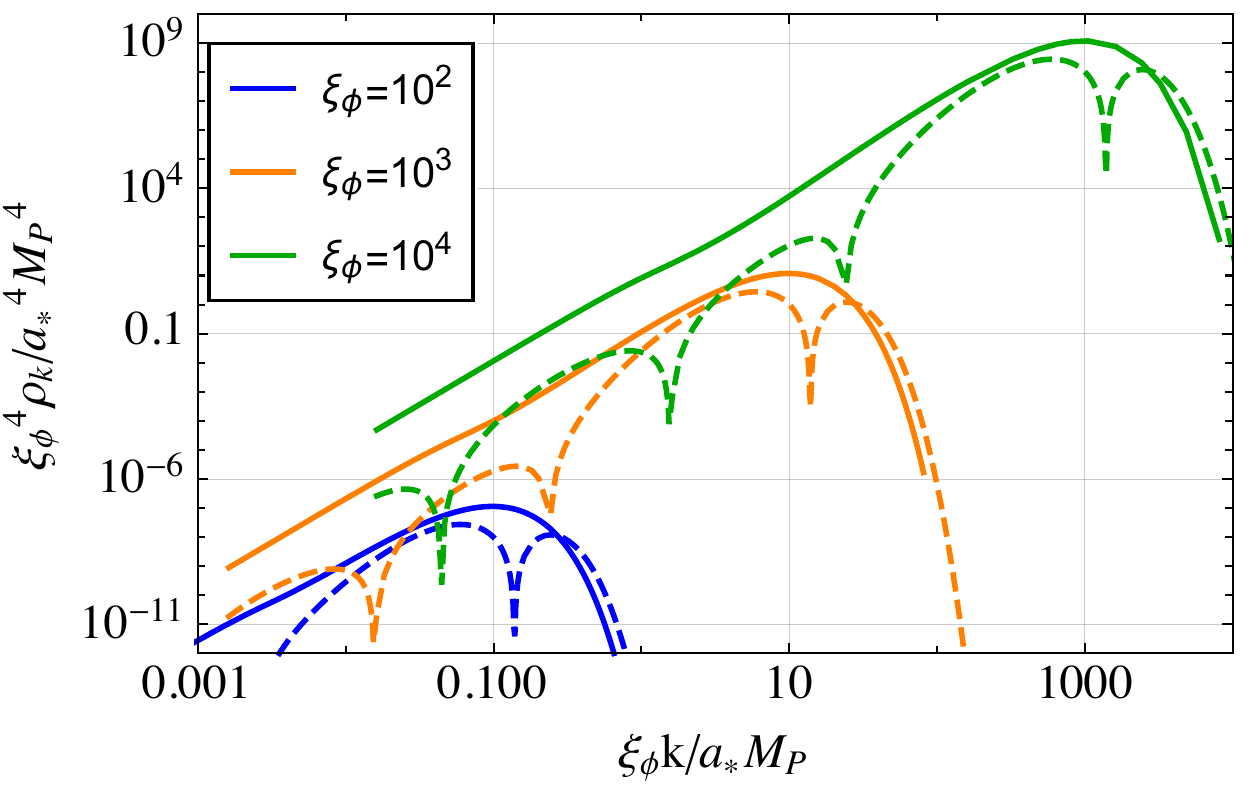}
	\caption{\small Particle number and energy density for Higgs inflation. Solid lines show the numerical solution of the mode equation and the dashed lines the results from the semiclassical approximation. \\ \emph{Left:} particle number density for Higgs inflation with $\xi_\phi =100$ and $\xi_\chi = 100,300,500$ (blue, orange, green respectively).  \\ \emph{Right:} energy density for Higgs inflation with $\xi_\phi = 10^2, 10^3, 10^4$ (blue, orange, green respectively) and $\xi_\chi = \xi_\phi$. }
	\label{fig:pp_Higgs_inf}
\end{figure}

We now study the spectrum of the produced particles in detail 
with the method developed in Sec.~\ref{sec:pp_field_space},
confirming the above estimation.
The relevant geometrical quantities are given by
\begin{align}
	{R^\chi}_{\phi\phi\chi} &= -\frac{1}{6M_P^2\Omega^4 \left(\Omega^2 + 6\xi_\phi^2\phi^2/M_P^2\right)}
	\left[
	\Omega^4\left(1+6\xi_\phi\right)\left(1+6\xi_\chi\right) - \left(\Omega^2 + \frac{6\xi_\phi^2 \phi^2}{M_P^2}\right)^2
	\right], 
	\label{eq:mass_curvature_HI} \\
	\nabla^\chi V_\chi &= \frac{\phi^2}{\Omega^4}\left[-\lambda_\phi \xi_\chi \frac{\phi^2}{M_P^2} + g^2 \Omega^2
	- \frac{\lambda_\phi \xi_\phi\left(1+6\xi_\chi\right) \phi^2/M_P^2}{\Omega^2 + 6\xi_\phi^2 \phi^2/M_P^2}
	\right],
	\label{eq:mass_pot_HI}
\end{align}
where $\chi$ is set to zero in $\Omega$ in these expressions,
and hence the effective mass is given by
\begin{align}
	m_\chi^2 &= \frac{\dot{\phi}^2}{6M_P^2\Omega^4 \left(\Omega^2 + 6\xi_\phi^2\phi^2/M_P^2\right)}
	\left[
	\Omega^4\left(1+6\xi_\phi\right)\left(1+6\xi_\chi\right) - \left(\Omega^2 + \frac{6\xi_\phi^2 \phi^2}{M_P^2}\right)^2
	\right] \nonumber \\
	&+ \frac{\phi^2}{\Omega^4}\left[-\lambda_\phi \xi_\chi \frac{\phi^2}{M_P^2} + g^2 \Omega^2
	- \frac{\lambda_\phi \xi_\phi\left(1+6\xi_\chi\right) \phi^2/M_P^2}{\Omega^2 + 6\xi_\phi^2 \phi^2/M_P^2}
	\right].
	\label{eq:chi_mass_Higgs_inf}
\end{align}
It is now numerically straightforward to solve Eq.~\eqref{eq:alpha_beta_tau}
with the background equations of motion
and obtain the spectrum~\eqref{eq:fchi_tau}.

In order to understand the numerical results analytically,
we focus on the leading order terms in the large $\xi_\phi$ and $\xi_\chi$ limit,
since they are expected to dominantly contribute to the particle production.
This simplifies the mass to
\begin{align}
	m_\chi^2 \simeq \frac{\xi_\chi}{\xi_\phi}
	\frac{6\xi_\phi^2}{M_P^2}\frac{\dot{\phi}^2}{\left(1+6\xi_\phi^2\phi^2/M_P^2\right)}.
\end{align}
If we ignore the Hubble expansion, the energy density is conserved,
which in our case leads to
\begin{align}
	\dot{\phi}^2
	\simeq
	\frac{\lambda_\phi}{2}\frac{\tilde{\Phi}^4}{1 + 6\xi_\phi^2 \phi^2/M_P^2},
	\quad
	\tilde{\Phi}^4 = \frac{\Phi^4}{\left(1 + \xi _\phi\Phi^2/M_P^2\right)^2},
\end{align}
where $\Phi = \mathcal{O}(M_P/\sqrt{\xi_\phi})$ is the inflaton amplitude at the end of inflation.
Here we again kept only the leading-order terms in $\xi_\phi$.
Now these equations correspond to the second example in Sec.~\ref{subsec:pia_examples},
and hence we can directly apply the analysis there.
As a result, we estimate the spectrum as
\begin{align}
	f_k \simeq \frac{\pi^2}{4}
	\abs{\exp\left[-2\sqrt{\frac{\xi_\chi}{\xi_\phi}}\int_0^{\bar{\phi}_+} d\bar{\phi}\, F \right]
	+ \exp\left[-2\sqrt{\frac{\xi_\chi}{\xi_\phi}}\int_0^{\bar{\phi}_-} d\bar{\phi}\, F \right]
	- 2\exp\left[-2\sqrt{\frac{\xi_\chi}{\xi_\phi}}\int_0^{1} d\bar{\phi}\, F \right]
	}^2,
	\label{eq:pia_Higgs_inf}
\end{align}
where 
\begin{align}
	\bar{\phi} = \frac{\xi_\phi \phi}{\sqrt{6} M_P},
	\quad
	\bar{k}^2 = \frac{\xi_\phi}{\xi_\chi}\frac{k^2 M_P^2/a_*^2}{3\lambda_\phi \xi_\phi^2 \tilde{\Phi}^4},
	\quad
	F = \sqrt{\bar{k}^2\left(1-\bar{\phi}^2\right) + \frac{1}{1-\bar{\phi}^2}},
	\quad
	\bar{\phi}_{\pm} = \sqrt{1\pm \frac{i}{\bar{k}}}.
\end{align}
We again take $\Phi$ as the inflaton field value at the point $\epsilon = 1$,
and use the scale factor at the first turning point $a_*$ in the definition of $\bar{k}$
in these expressions.

In Fig.~\ref{fig:pp_Higgs_inf}, we compare the numerical results with the analytical estimate~\eqref{eq:pia_Higgs_inf}. We take
\begin{align}
	g^2 = \lambda_\phi \left(\frac{\xi_\chi}{\xi_\phi}+ 0.2 \right)\,,
\end{align}
and fix $\lambda_\phi$ by Eq.~\eqref{eq:Higgs_inf_CMB}.
 We again initialize the background field such that it follows the attractor solution in the region where we solve the mode equation. We initialize the mode functions 2 $e$-folds before the end of inflation and we evaluate the particle number and energy density at $\phi_*$.

The left panel of Fig.~\ref{fig:pp_Higgs_inf} shows the particle number spectrum for $\xi_\phi =100$ and $\xi_\chi = 100, 300, 500$. The physical momenta have been rescaled by $M_P/\xi_\phi$, which roughly corresponds to the UV cut-off scale, which scales as $\Lambda \sim M_P/\sqrt{\xi_\phi\xi_\chi}$.  For $\xi_\chi = 300, 500$ the particle number is smaller than 1 and the Born approximation is expected to be valid. The semiclassical approximation captures the UV part of the spectrum very well, even in the case where we include the expansion of the universe. For $\xi_\chi=100$ the produced particle number is larger than 1 and the Born approximation is thus less powerful in the IR. Yet, the UV-tail is practically the most relevant, as can be seen in the right panel of Fig.~\ref{fig:pp_Higgs_inf}, where the contribution to the energy density is shown for $\xi_\phi=10^2, 10^3, 10^4$ and $\xi_\chi=\xi_\phi$. Even in this case, where the Born approximation breaks down in the IR, as the particle number exceeds 1 for all three cases, the approximation works well at the peak of the energy density. The value of $(k/a_*)_\text{max}$ is reproduced up to a factor 2. This also becomes clear in the right panel of Fig.~\ref{fig:kmax}, where the ratio of $(k/a_*)_\text{max}$ over the cut-off scale (taken to be $M_P/\xi_\phi$) is plotted as a function of $\xi_\phi$. According to our numerical solution, unitarity is violated for $\xi_\phi \gtrsim 300$ and this value is reasonably reproduced by the semiclassical result.
We also emphasize that the semiclassical formula works well even for $\xi_\phi = 10^2$, for which
unitarity is not violated.

Let us make a brief comment on possible UV completions of Higgs inflation.
For this case, some UV completions have been suggested in the literature~\cite{Lerner:2010mq,Giudice:2010ka,Barbon:2015fla,Ema:2017rqn,Lee:2018esk,Gorbunov:2018llf}. 
In the large-$N$ limit, a healing mechanism arises~\cite{Aydemir:2012nz,Calmet:2013hia,Ema:2019fdd}, which restores unitarity,
where $N$ denotes the number of real scalar fields ($N=4$ for the Standard Model Higgs).
This healing mechanism is now understood in the language of the (frame-independent) non-linear sigma model
with the scalaron corresponding to the $\sigma$-meson~\cite{Ema:2020zvg}.
 These UV completions can greatly alter the preheating dynamics, as was demonstrated in Refs.~\cite{He:2018mgb,Bezrukov:2019ylq,He:2020ivk,Bezrukov:2020txg,Hamada:2020kuy}.

\subsubsection*{$\alpha$-attractor inflation}

Next we consider $\alpha$-attractor inflation.
We consider the following action
\begin{align}
	S &= \int d^4x \sqrt{-g}\left[\frac{M_P^2}{2}R 
	+ \frac{\left(\partial \phi^a\right)^2}{2\left(1- \phi^a \phi^a/\Lambda^2\right)^2}
	- \frac{\lambda}{4}\left(\phi^a \phi^a\right)^2
	 \right],
\end{align}
where $\phi^a = \phi, \chi$.
This complex scalar field version of $\alpha$-attractor inflation is discussed, \emph{e.g.}, in Ref.~\cite{Krajewski:2018moi}.

We now see that this model does not violate unitarity during preheating
by using the condition~\eqref{eq:unitarity_condition}.
In this model, inflation happens when the inflaton field is close to the pole of its kinetic term, $\phi \sim \Lambda$,
and hence the height of the inflaton potential at the end of inflation is given by
\begin{align}
	V(\Phi) \sim \lambda \Lambda^4.
\end{align}
Since the cut-off scale of this theory is of order $\Lambda$, we obtain
\begin{align}
	\frac{V(\Phi)}{\Lambda^4} \sim \lambda \lesssim \mathcal{O}(1),
\end{align}
where we assume that $\lambda$ is less than order one by perturbativity constraints.
Then, according to the discussion in Sec.~\ref{subsec:unitarity},
we expect that unitarity is preserved during preheating in this model.
Indeed no unitarity violation during preheating is observed in the literature.

In this model, the geometrical quantities are given by
\begin{align}
	{R^{\chi}}_{\phi\phi\chi} &= \frac{4}{\Lambda^2}\frac{1}{\left(1-\phi^2/\Lambda^2\right)^2},
	\quad
	\nabla^\chi V_\chi = \lambda \phi^2 \left(1 - \frac{\phi^4}{\Lambda^4}\right),
\end{align}
and hence the mass term is given by
\begin{align}
	m_\chi^2 = -\frac{4}{\Lambda^2}\frac{\dot{\phi}^2}{\left(1-\phi^2/\Lambda^2\right)^2}
	+\lambda \phi^2 \left(1 - \frac{\phi^4}{\Lambda^4}\right).
\end{align}
In particular, the contribution from the target space curvature is tachyonic.
This requires a separate treatment and we cannot straightforwardly 
apply our semiclassical method in Sec.~\ref{sec:pp_field_space}.
Therefore we do not discuss particle production of this model any further.
See Refs.~\cite{Krajewski:2018moi,Iarygina:2018kee,Iarygina:2020dwe} 
for more on the preheating dynamics of this model.

\subsubsection*{Higgs-Palatini inflation}

Finally we consider Higgs-Palatini inflation.
The action in the Jordan frame is given by
\begin{align}
	S = \int d^4 x \sqrt{-g_J} \left [ \frac{M_P^2}{2}\Omega^2 R_J(\Gamma)
	+ \frac{1}{2} g_J^{\mu\nu} \partial_\mu \phi^a \partial_\nu \phi^a 
	- \frac{\lambda}{4}\left(\phi^a\phi^a\right)^2 \right]\, ,
\end{align}
where the subscripts ``$J$" again denote quantities in the Jordan frame, $\phi^a = \phi, \chi$, and we defined 
\begin{align}
	\Omega^2 = 1 + \frac{\xi \phi^a \phi^a}{M_P^2}.
\end{align}
This form of the action is the same as Eq.~\eqref{eq:action_Higgs_inf},
but an essential point of the Palatini formulation is that
one takes the spin connection independent of the vierbein
at the level of the action, 
and it is given as a solution of the equation of motion (which results in a constraint equation).
Since the equation of motion is of first order in derivatives,
it is sometimes called the first order formalism~\cite{Freedman:2012zz}.

It is again convenient to go to the Einstein frame by rescaling the metric as
\begin{align}
	g_{J\mu\nu} \rightarrow g_{\mu\nu} = \Omega^2 g_{J\mu\nu},
\end{align}
resulting in
\begin{align}
	S = \int d^4 x \sqrt{-g} \left[\frac{M_P^2}{2} R + \frac{1}{2\Omega^2} g^{\mu\nu} \partial_\mu \phi^a \partial_\mu \phi^a 
	-\frac{\lambda}{4}\frac{\left(\phi^a\phi^a\right)^2}{\Omega^4} \right].
\end{align}
Note that the target space metric is different from that of Higgs inflation
because the Ricci scalar transforms differently under the Weyl transformation
in the metric and Palatini formalisms.
In the Einstein frame, the equation of motion tells us that the spin connection
is given in terms of the vierbein in the standard form,
and hence we can equally think of it as the metric theory.

We now apply the condition~\eqref{eq:unitarity_condition} to this model.
In this model, inflation happens for $\phi \gtrsim M_P/\sqrt{\xi}$
where the additional term in the target space metric becomes important,
and hence the height of the inflaton potential at the end of inflation is given by
\begin{align}
	V(\Phi) \sim \frac{\lambda M_P^4}{\xi^2}.
\end{align}
Since the cut-off scale of this theory is $\Lambda \sim M_P/\sqrt{\xi}$, we obtain
\begin{align}
	\frac{V(\Phi)}{\Lambda^4} \sim \lambda \lesssim \mathcal{O}(1),
\end{align}
where we again assume that $\lambda$ is less than order one, from the perturbativity requirement.
Thus, according to the discussion in Sec.~\ref{subsec:unitarity},
we expect that unitarity is preserved during preheating in this model.
This is in contrast to Higgs inflation in the metric formalism,
where the ratio can be greater than unity for large enough $\xi$ (see Eq.~\eqref{eq:unitarity_Higgs_inf}).
An essential difference is that the height of the inflaton potential and the cut-off scale depend on
the different scales, $M_P/\sqrt{\xi}$ and $M_P/\xi$, in Higgs inflation, 
while they depend on the common scale $M_P/\sqrt{\xi}$ in Higgs-Palatini inflation.
Indeed no unitarity violation during preheating is observed in the literature.

In this model, the inflaton amplitude at the end of inflation is given by $\Phi \sim (M_P/\sqrt{\xi}) \ln \xi$~\cite{Bauer:2008zj}.
Since the CMB requires $\xi \sim 10^{10} \lambda$, 
the logarithmic term in $\Phi$ can be sizeable, $\ln \xi \sim \mathcal{O}(10)$, unless $\lambda$ is tiny.
This causes a scale separation between $\Phi$ and $\Lambda \sim M_P/\sqrt{\xi}$ 
that may justify the approximation $\Phi \rightarrow \infty$ in Eq.~\eqref{eq:born_general}.
Since the effective mass term and the conserved quantities are not in the form discussed 
in Sec.~\ref{subsec:pia_examples},
we need a separate analysis in this model. Therefore, although it would be interesting,
we leave a detailed semiclassical study of this model as a future work.
See also Ref.~\cite{Rubio:2019ypq} for more on the preheating dynamics of this model.

\subsubsection*{Comments on the unitarity issue of Higgs inflation}
Before closing this subsection, let us clarify our attitude on the unitarity issue
of (metric) Higgs inflation in this paper, since different points of view are taken on this issue in the literature.

It is well-known that the cut-off scale of Higgs inflation for large $\xi$ is given by $M_P/\xi~(\ll M_P)$
in the vacuum~\cite{Burgess:2009ea,Barbon:2009ya,Burgess:2010zq,Hertzberg:2010dc,Bezrukov:2010jz}, 
referred to as the unitarity issue of Higgs inflation.
Ref.~\cite{Bezrukov:2010jz} pointed out that the cut-off scale increases 
as the Higgs gets a large vacuum expectation value,
and hence the unitarity issue does not necessarily spoil the inflationary prediction of Higgs inflation.
However, we are aware of two types of arguments that cast doubt 
on the validity of Higgs inflation, besides unitarity violation during preheating.

Ref.~\cite{Barbon:2015fla} constructed an explicit UV completion of Higgs inflation
by adding an additional scalar field to the theory.
The authors found that it is the additional scalar field, and not the Higgs, that dominantly drives inflation,
and hence the model ceases to be ``Higgs inflation" after the UV completion.
They expect that this situation is generic and the inflationary prediction in general depends 
on the specific form of UV completion.

Refs.~\cite{Burgess:2014lza,Fumagalli:2016lls} take a slightly different point of view.
The main concern of the authors is on the relation between the parameters such as $\lambda$,
or more generally the shape of the Higgs potential, at the electroweak scale and the inflationary scale.
According to their argument, even if one assumes the existence of a UV theory 
in which inflation is described by the Higgs field only (this itself may be a strong assumption based on Ref.~\cite{Barbon:2015fla}),
one has to introduce a ``threshold correction" that parametrizes 
one's ignorance of the connection between low and high energy physics.
This prohibits one to relate the parameters at the electroweak scale with the inflationary prediction, to say the least.
The threshold correction may even push $\lambda$ negative, making Higgs inflation impossible.

In our discussion of unitarity violation of Higgs inflation during preheating above, 
we simply ignore these arguments,
\emph{e.g.}, by assuming that a UV completion does not alter the inflationary dynamics of the Higgs,
and taking $\lambda$ positive without caring about its relation to the value at the electroweak scale.
Our point is that, even if we put aside the points raised here, Higgs inflation still causes an issue during preheating, 
and an explicit UV completion is 
anyway unavoidable for $\xi \gtrsim \mathcal{O}(100)$.
In other words, we have served yet another argument that demands an explicit UV completion of Higgs inflation.
Although not widely discussed in the literature,
the same line of argument should apply to running kinetic inflation.

\subsection{Application to scalar dynamics other than inflaton}
\label{subsec:saxion}

So far we have considered particle production induced by the inflaton motion.
In theories beyond the Standard Model, there often appear scalar fields other than the inflaton
that exhibit coherent motion in the field space and lead to particle production in the early universe.
In supergravity, for example, the scalar potential often has many flat directions along which the potential energy is zero in the supersymmetric limit. 
Such flat directions may be called moduli and their impacts significantly depend on their masses and interaction strengths.

Here we consider one explicit example, the dynamics of the saxion in supergravity~\cite{Kim:1983ia,Rajagopal:1990yx,Kim:1992eu,Lyth:1993zw,Chun:1995hc,Choi:1996vz,Kasuya:1996ns,Asaka:1998xa,Kawasaki:2007mk,Moroi:2013tea,Ema:2017krp}. The saxion is a scalar partner of the axion that solves the strong CP problem in quantum chromodynamics~\cite{Peccei:1977hh,Kim:1986ax}.
The flatness of the saxion potential is ensured by the holomorphy of the superpotential combined with the Peccei-Quinn (PQ) global U(1) symmetry in the supersymmetric limit. 
The action of a supersymmetric axion model is given by
\begin{align}
	&S=\int d^4x \sqrt{-g}\left[\frac{M_P^2}{2}R +|\partial\psi_1|^2+|\partial\psi_2|^2-V(\psi_1,\psi_2) \right],  \label{S_saxion}\\
	&V(\psi_1,\psi_2)=\lambda^2\left| \psi_1 \psi_2-\Lambda^2 \right|^2 + m_1^2|\psi_1|^2+m_2^2|\psi_2|^2,
\end{align}
where $\psi_1$ and $\psi_2$ are complex scalars and we assume $m_1^2\simeq m_2^2 (\equiv m^2) \ll \Lambda^2$. The action is invariant under the global U(1) symmetry, $\psi_1 \to e^{i\theta}\psi_1$ and $\psi_2 \to e^{-i\theta}\psi_2$ with a real parameter $\theta$. 
Among four real degrees of freedom, two obtain masses of order $\Lambda$, one obtains a mass of order $m$ corresponding to the saxion, and the last one is massless, which is regarded as the axion.

Let us suppose that we can integrate out the heavy modes of mass $\sim\Lambda$ to consider the saxion dynamics. By using the constraint $\psi_1\psi_2=\Lambda^2$ and decomposing $\psi_1$ as $\psi_1=(\phi/\sqrt{2})e^{i\chi/\phi}$, we obtain the effective action for the saxion $\phi$ and axion $\chi$ as 
\begin{align}
	S =\int d^4x \sqrt{-g}\left[\frac{M_P^2}{2}R + \frac{1}{2}h_{ab} g^{\mu\nu}\partial_\mu\phi^a\partial_\nu\phi^b - V(\phi)\right],
\end{align}
where we defined the vector $\phi^a=(\phi,\chi)$ and
\begin{align}
	h_{ab}=\left(1+\frac{4\Lambda^4}{\phi^4}\right)
	\begin{pmatrix}
		\displaystyle 1+\frac{\chi^2}{\phi^2} &\displaystyle -\frac{\chi}{\phi} \\
		\displaystyle -\frac{\chi}{\phi} & \displaystyle 1
	\end{pmatrix},
	\quad
	V(\phi)= \frac{m^2}{2}\left(1+\frac{4\Lambda^4}{\phi^4}\right) \phi^2.
\end{align}
Thus we obtained a curved target space with typical curvature scale $\Lambda$ as an induced metric.
It has been pointed out in Ref.~\cite{Ema:2017krp} that the effective mass of the axion in this setup shows a spiky behavior and the axion particle production from the saxion dynamics is very different from the standard scenario.

The general arguments in Secs.~\ref{subsec:mass_curvature} and~\ref{subsec:unitarity} can be equally applied to this case.
The relevant geometrical quantities are given by
\begin{align}
	{R^\chi}_{\phi\phi\chi} = \frac{2\phi^2}{\Lambda^4}\frac{1}{\left(1+\phi^4/4\Lambda^4\right)^2},
	\quad
	\nabla^\chi V_\chi = m^2\left(\frac{1-\phi^4/4\Lambda^4}{1+\phi^4/4\Lambda^4}\right)^2,
\end{align}
and hence the effective mass of the canonical axion field is given by
\begin{align}
	m_\chi^2 =
	-\frac{2\phi^2}{\Lambda^4}\frac{ \dot{\phi}^2}{\left(1+\phi^4/4\Lambda^4\right)^2}
	+ m^2\left(\frac{1-\phi^4/4\Lambda^4}{1+\phi^4/4\Lambda^4}\right)^2.
\end{align}
The first term that originates from the target space curvature 
shows the spiky behavior around $\abs{\phi}\lesssim \Lambda$.
Assuming that the initial saxion field value satisfies $\Phi \gg \Lambda$, the typical energy scale of the produced particles 
is given by 
\begin{align}
	\left(\frac{k}{a_*}\right)_\mathrm{max}^2 \sim \frac{V(\Phi)}{\Lambda^2},
\end{align}
where $a_*$ is the scale factor at the first turning point of the saxion.
If we require that it should be below the cutoff scale $\Lambda$, we again obtain the same condition as in the preheating analysis, $V(\Phi) \lesssim \Lambda^4$. 
If this condition is violated, integrating out the heavy fields and the use of the effective action may not be justified. 
For a large enough initial field value, $\Phi\gtrsim \Lambda^2/m$, we must use the original renormalizable action~(\ref{S_saxion}) for the analysis of particle production. Actually for such a large initial saxion field value, the field trajectory significantly deviates from the flat direction and passes through the region $\psi_1=0$ or $\psi_2=0$, which may lead to nonthermal symmetry restoration and formation of topological defects~\cite{Kofman:1995fi,Moroi:2013tea,Kawasaki:2013iha}. Thus the cosmological effects of the saxion dynamics are completely different depending on whether the condition $V(\Phi) \lesssim \Lambda^4$ is satisfied or not.

This particular example demonstrates the phenomenological importance of the unitarity violation condition
beyond the context of inflation.
Once the low energy description breaks down, one must go back to a UV completed model. 
In the UV model, however, the dynamics should include heavy modes and they can have drastic effects on cosmology. The formation of topological defects, as seen in the saxion case, is an example of possible dangerous effects. Another possible dangerous effect is production of heavy particles that are stable against decay or charged under some (approximate) symmetry. 

Here is a general remark on the moduli cosmology.
In the context of string theory, the size and geometry of the compactified extra dimensions may be characterized by the moduli fields from the four-dimensional viewpoint, and their interactions are often suppressed by some high energy scale, leading to the notorious cosmological moduli problem~\cite{Banks:1993en,Coughlan:1983ci}. Usually the moduli potential is assumed to be a simple quadratic form, but in general the flat directions may have a complicated structure in field space. In such a case particle production by the moduli and its cosmological consequences can be much different from the naive analysis.

\section{Summary}
\label{sec:summary}

In this paper, we have studied particle production and unitarity violation in inflationary models with multiple scalar fields and a curved target space during preheating after inflation. In these models the curvature of the target space causes a spiky feature in the mass of the other scalar particle(s) that couple to the inflaton. This feature results in efficient particle production and possibly in unitarity violation.

In Sec.~\ref{sec:pp_field_space} we demonstrate how the particle number spectrum after the first burst of particle production can be estimated in a semiclassical analysis, relying on the Born approximation ($|\alpha_k|^2 \sim 1, |\beta_k|^2 \ll 1$). Instead of $t$, we use the inflaton field $\phi$ as the time variable,
hence referring to it as an analysis in (scalar) field space. 
This is possible since the inflaton field is a monotonic function of time between the end of inflation and its first turning point.
We replace the finite integration range, from $\Phi$ to $-\Phi$ to an integration over the entire $\phi$-axis,
where $\Phi$ is the inflaton  oscillation amplitude. Upon using Cauchy's residue formula, the spectrum can then be expressed as a sum over poles. We demonstrate this procedure for two example mass terms, with poles in $\phi$ of order one and two, by making use of the potential energy at the end of inflation as a conserved quantity. The semiclassical formula for the occupation number shows an excellent agreement with the numerical results for the two given examples.

In Sec.~\ref{sec:specmodels} the analytical expressions are put to the test for two inflationary models: running kinetic inflation and Higgs inflation, which are similar to the pole of order one and two cases respectively. 
In the case of running kinetic inflation, the semiclassical approximation gives an excellent estimate of the momentum $k_\text{max}$ at the peak of the produced energy density spectrum, which is most relevant for the question whether unitarity is violated. The semiclassical approximation also gives a good description of the UV-part of the spectrum in the case of Higgs inflation, where it allows a determination of $k_\text{max}$ up to a factor 2.

In Sec.~\ref{subsec:unitarity} we provide a simple criterion for unitarity violation during reheating
\begin{align*}
	V(\Phi) \gtrsim \Lambda^4\, ,
\end{align*}
that is, unitarity is violated if the inflaton potential at the end of inflation $V(\Phi)$ is larger than the fourth power of $\Lambda$, where $\Lambda$ is the curvature of the target space as well as the UV cut-off scale of the theory. The criterion immediately indicates that, for couplings consistent with the CMB normalization, unitarity can be violated for running kinetic inflation and Higgs inflation. This is also confirmed in our semiclassical analysis and numerical computations in Sec.~\ref{sec:specmodels}, confirming that unitarity is violated for $\Lambda \lesssim 10^{-2} M_P$ for running kinetic inflation and for $\xi_\phi \gtrsim 100$ for Higgs inflation. According to our condition, unitarity is not violated for $\alpha$-attractor models and Higgs-Palatini inflation, as is consistent with the literature~\cite{Krajewski:2018moi,Iarygina:2018kee,Iarygina:2020dwe, Rubio:2019ypq}.

Furthermore, we point out that particle production due to a curved target space is not exclusive to the preheating phase. As an example, we show in Sec.~\ref{subsec:saxion} how integrating out the heavy field in a supersymmetric axion model results in an effective theory for the axion and the saxion with a curved target space. The axion has a spike-like mass and unitarity violation is again possible, implying that the effective theory is not valid. The case of unitarity violation actually corresponds to the formation of topological defects in the original theory. 

Unitarity violation is of great phenomenological importance, as it calls for a UV completion of the theory. 
If one starts from a UV-complete theory, as in the case of the saxion,
one can simply go back to the original UV-complete theory. 
If one starts from low energy theories as in the cases of running kinetic inflation and Higgs inflation, however,
this is a nontrivial requirement. We discuss several implications of the UV completion of running kinetic inflation and Higgs inflation   in Sec.~\ref{sec:specmodels}.
For running kinetic inflation, 
it is shown that phenomenological consequences, possibly including domain wall formation, crucially depend on the concrete model of UV completion.
This explicitly shows the limit of using effective theory during (p)reheating in the presence of unitarity violation.

\section*{Acknowledgements}
This work was partly funded by the Deutsche Forschungsgemeinschaft under Germany's Excellence Strategy - EXC 2121 ``Quantum Universe'' - 390833306.
The work of RJ was supported by Grants-in-Aid for JSPS Overseas Research Fellow (No. 201960698).
This work was also supported by JSPS KAKENHI Grant (Nos. 18K03609 [KN] and 17H06359 [KN]). We thank Marieke Postma for useful comments.

\small
\bibliographystyle{utphys}
\bibliography{ref}
  
\end{document}